\documentclass{raa_twocolumn}            

\usepackage{graphicx,times}             
\usepackage{natbib}
\usepackage{amssymb,amsmath}
\usepackage{comment}
\usepackage{float}
\bibpunct{(}{)}{;}{a}{}{,}
\usepackage[pagebackref=true]{hyperref}

\begin{document}

  \title{Early results from the SVOM Observatory Science program}

   \volnopage{Vol.0 (202x) No.0, 000--000}      
   \setcounter{page}{1}          

   \author{A. Coleiro  
      \inst{1,*}\footnotetext{$*$Corresponding Author}
   \and L. Tao
      \inst{2}
      \and F. Cangemi
      \inst{1}
      \and X. Han
      \inst{3}
      \and M. Brunet
      \inst{4}
      \and N. Dagoneau
      \inst{5}
      \and A. Foisseau
      \inst{1}
      \and A. Goldwurm
      \inst{1, 5}
      \and S. Guillot
      \inst{4}
      \and N. Jiang
      \inst{6}
      \and C. Lachaud
       \inst{1}
      \and S. Le Stum
      \inst{1}
      \and P. Maggi
      \inst{7}
      \and D. Rawat
      \inst{7}
      \and J. Rodriguez
      \inst{8}
      \and C.W. Wang
      \inst{2}
      \and J. Wang
      \inst{3}
      \and W. Xie
      \inst{3}
      \and L. Zhang
      \inst{2}
      \and L. Bouchet 
      \inst{4}
          \and M. Clavel 
      \inst{9} 
       \and
       Z. Feng
       \inst{10}
        \and O. Godet 
      \inst{4}
      \and D. Götz
      \inst{5}
      \and
      D. Li
      \inst{10}
      \and Z. Li
      \inst{11} 
      \and L. Lin
      \inst{12} 
      \and T. Maiolino
      \inst{13} 
      \and J. Wu 
      \inst{14} 
      \and L. Xin
      \inst{3}
    \and S.L. Xiong 
      \inst{2}
      \and X. Xu 
      \inst{15} 
       \and J. Zhang
      \inst{16} 
      \and S.J. Zheng
      \inst{2}
      \and B. Cordier
\inst{5}
\and J. Wei
\inst{3}
\and S. Basa
\inst{17} 
\and A. Claret
\inst{5}
\and F. Daigne
\inst{18} 
\and J. Deng
\inst{3}
\and Y.W. Dong
\inst{2}
\and E. W. Liang 
\inst{19} 
\and Y. Qiu
\inst{3}
\and S. Vergani
\inst{20} 
\and C. Wu
\inst{3}
\and B. Zhang
\inst{21} 
\and S. N. Zhang
\inst{2}
 on behalf of the SVOM collaboration
  \inst{22}
  }

   \institute{Universit\'e Paris Cit\'e, CNRS, Astroparticule et Cosmologie, F-75013 Paris, France; {\it coleiro@apc.in2p3.fr}\\
   \and
             State Key Laboratory of Particle Astrophysics, Institute of High Energy Physics, Chinese Academy of Sciences, Beijing 100049, P. R. China\\
\and
National Astronomical Observatories, Chinese Academy of Sciences,
             Beijing 100101, China\\
\and
IRAP, Universit\'e de Toulouse, CNRS, CNES, Toulouse, France\\
        \and
             CEA Paris-Saclay, IRFU/DAp-AIM, 91191 Gif sur Yvette, France\\
        \and
 Department of Astronomy, University of Science and Technology of China, Hefei 230026, People’s Republic of China\\
 \and
 Observatoire Astronomique de Strasbourg, Université de Strasbourg, CNRS, 11 rue de l’Université, F-67000 Strasbourg, France\\
 \and
 Université Paris-Saclay, Université Paris Cité, CEA, CNRS, AIM, 91191 Gif-sur-Yvette, France\\
 \and
  Univ. Grenoble Alpes, CNRS, IPAG, 38000 Grenoble, France\\
  \and
  Laboratory of Satellite Mission Operations, National Space Science Center, Chinese Academy of Sciences, Beijing 100190, China\\
  \and
    School of Science, Qingdao University of Technology, Qingdao 266525, P.R. China\\
  \and
  Department of Astronomy, Beijing Normal University, Beijing 100875, People’s Republic of China\\
  \and
  Laboratoire Univers et Particules de Montpellier, Université Montpellier, CNRS/IN2P3, F-34095 Montpellier, France\\
  \and
  Department of Astronomy, Xiamen University, Xiamen, Fujian 361005, People’s Republic of China
  \and
School of Astronomy and Space Science, Nanjing University, 210023 Nanjing, Jiangsu, China\\
\and
  School of Physics, Beijing Institute of Technology, Beijing 100081, People’s Republic of China\\
  \and  
  Aix Marseille Universit\'e., CNRS, CNES, LAM, Marseille, France\\
  \and 
  Sorbonne Universit\'e, CNRS, UMR 7095, Institut d'Astrophysique de Paris, 98 bis bd Arago, F-75014 Paris, France\\
\and
Guanxi Key Laboratory for Relativitic Astrophysics, Guangxi University, Nanning 530004, P. R. China\\
\and 
  LUX, Observatoire de Paris, Universit\'e PSL, CNRS, Sorbonne Universit\'e, Meudon, 92190, France\\
  \and
Hong Kong Institute for Astronomy and Astrophysics, University of Hong Kong, Pokfulam Road, P. R. China\\
\and
\url{https://fsc.svom.org/home/collaboration/collaborators}\\
\vs\no
   {\small Received 202x month day; accepted 202x month day}}

\abstract{ We present the organisation and early results from the Observatory Science program of the Space-based multi-band astronomical Variable Objects Monitor (SVOM), based on data collected between July 2024 and December 2025. Although primarily designed for gamma-ray burst studies, SVOM’s wide-field, multi-wavelength instruments enable a broad range of high-energy astrophysical investigations. We summarize the execution and performance of the General Program and Target-of-Opportunity observations, and we describe the frameworks used for serendipitous source detection and monitoring with the ECLAIRs coded-mask instrument. Over this period, SVOM carried out more than a thousand pointed observations and detected several hundred non-GRB high-energy sources, mainly X-ray binaries, as well as blazars, stellar flares, magnetars, and unidentified events. We highlight some key results, including the monitoring of the microquasar Cygnus X-1, the detection of burst oscillations from the Low-Mass X-ray Binary 4U~0614+091, the spectral-state monitoring of Aql~X-1, the first SVOM detection of an X-ray blazar flare from 1ES~1959+650, and observations of a stellar flare from HD~22468. These results demonstrate SVOM’s strong capabilities for time-domain astrophysics beyond its core GRB program.
}

   \authorrunning{A. Coleiro, L. Tao et al.}            
   \titlerunning{SVOM Observatory Science Program: early results}  

   \maketitle

%
%
\section{Introduction}           
\label{sect:intro}

Launched on June 22nd 2024, the SVOM mission is dedicated to the study of high-energy transient sources across multiple wavelengths \citep{Cordier+etal+2026}. After a six-month commissioning and validation phase, scientific operations formally began in January 2025. The mission is planned to operate for a nominal duration of three years, followed by a two-year extended phase. The SVOM payload carries four complementary instruments: ECLAIRs \citep{Godet+etal+2026}, a wide-field coded-mask hard X-ray (4--150\,keV) imager; GRM \citep{Sun+etal+2026}, a gamma-ray monitor extending the energy coverage of ECLAIRs up to 5\,MeV; MXT \citep{Götz+etal+2026}, a 0.2--10\,keV microchannel X-ray telescope for precise X-ray localization and follow-up; and VT \citep{Qiu+etal+2026}, a 40-cm diameter optical telescope operating simultaneously in blue and red channels. Although its primary objective is the detection and follow-up of gamma-ray bursts (GRBs; \citealt{Daigne+etal+2026}), the onboard instruments and ground segment are also well suited for the detection, monitoring, and characterization of a wide range of high-energy transients and variable sources.
Within this framework, SVOM’s scientific activities are organized into two major domains: GRB Science, focused on the study of GRBs and related phenomena, and Observatory Science, which encompasses all other astrophysical sources not directly associated with GRBs.

The Observatory Science comprises two complementary components: (i) an operational component dedicated to the detection of serendipitous high-energy sources, and (ii) a scientific component  structured into working groups that focus on specific classes of astrophysical sources. These groups oversee the analysis and interpretation of SVOM data, as well as the dissemination of scientific results.

The SVOM data supporting these activities may originate from any of the observing programs of the mission: the General Program, dedicated to pre-scheduled pointed observations; the Core Program which consists in observations triggered onboard following a transient detection by the ECLAIRs instrument; or the Target-of-Opportunity (ToO) Program, enabling rapid follow-up of transient astrophysical events identified through external alerts \citep{Cordier+etal+2026}.

In this article, we present the organisation and some results of the Observatory Science program obtained during the first year of SVOM’s scientific operations. Section~2 describes the General Program scheduling and provides an overview of the observations conducted in 2025 as well as preliminary scientific results from this program. Section~3 details the search, detection, and monitoring of non-GRB sources within the field-of-view (FoV) of the ECLAIRs coded-mask instrument. Section~4 highlights selected results related to SVOM ToO observations of external alerts. Section~5 summarizes the main findings and discusses future perspectives, and Section~6 concludes the article.

\section{SVOM General Program pointed observations}
\label{sect:GP_obs}

The SVOM General Program (GP) is dedicated to pre-planned pointed observations and complements the Core Program, which focuses on GRB science and Target-of-Opportunity (ToO) observations. The GP is designed in compliance with the system requirements of the GRB program, in particular the pointing strategy that optimizes ground-based follow-up of SVOM-detected GRBs (see Section \ref{sec:GP_description}). To implement this program, an annual Call for Proposals (CfP) is issued each year, inviting SVOM co-investigators and affiliated scientists to submit observation proposals on dedicated tools \citep{Han+etal+2026, Claret+etal+2026}, which are evaluated by an internal Time Allocation Committee (TAC). The CfP is also open to external researchers, provided they collaborate with a SVOM co-investigator.

\subsection{General Program Catalog, Scheduling and Observing Strategy}\label{sec:GP_description}

The GP planning framework structures the long-term  allocation of observational time within the SVOM mission. It is based on a GP catalog that aggregates all user observation requests approved by the TAC. This catalog, built based on the information (list of sources, observational strategy, etc.) provided by the users answering the CfP \citep{Han+etal+2026, Claret+etal+2026}, constitutes the basis for generating the global observation plan, which is refined into weekly schedules uploaded to the spacecraft. During the first year of SVOM scientific operations, the GP catalog encompassed all major objectives of the SVOM Observatory Science Program, including observations of Active Galactic Nuclei (AGNs), X-ray binaries, flaring stars, and other variable or transient astrophysical phenomena, as well as survey programs targeting, for example, the Virgo Cluster and the Small Magellanic Cloud. As weekly plans are executed on board, observations that are partially completed or interrupted (typically due to GRB triggers or Target-of-Opportunity overrides) are reinjected into the planning cycle.

A central element of GP planning is the implementation of the B1 reference pointing law (hereafter referred as B1 law), which requires the spacecraft to observe preferentially at $\sim$45$^\circ$ from the anti-solar direction and avoiding the Galactic plane to maximize GRB detection efficiency. Observations of GP targets may therefore be constrained to periods when the daily B1 law pointing is sufficiently close to the target direction, while unconstrained targets are placed within their visibility windows. When no higher-priority target satisfies the pointing constraints, backup targets (referred to as “fill-in sources”) are scheduled to maintain the productivity of the SVOM payload.
Scheduling also relies on the General Slew Point (GSP), a reference attitude derived from the subsolar position, which defines the start of each orbital visibility window. Observation durations are converted into an integer number of orbits based on the mean fraction of each orbit available for science, accounting for Earth occultation, the South Atlantic Anomaly (SAA) crossings, slews, the pointing stability, and potential interruptions by higher-priority GRB or ToO observations. 
The observation planning algorithm follows a structured multi-criteria framework that incorporates target priority (A+, A, B+, B, C), compliance with the B1 pointing requirement (which mandates that 85\% of GP observing time is spent on sources lying at less than $\pm10^\circ$ from the B1 law), the balance of observing time allocated to China- and France-based teams, and whether observations belonging to the same program have already been initiated. The scheduling algorithm iteratively inserts the GP target that best satisfies these constraints while minimizing global metrics that quantify the efficiency of visibility-time usage and the temporal dispersion of high-priority observations. For time slots where no catalogued sources can be scheduled, predefined pointings that comply with the B1 attitude law are used, even if they do not target any astrophysical sources.

This approach ensures the efficient use of observing time and maximizes the scientific return of pointed GP observations, while fully accommodating the demanding environment of a mission devoted to transient and time-domain astronomy. Despite frequent onboard GRB triggers and the high priority assigned to ToO observations, the planning framework remains robust, enabling the GP program to deliver meaningful scientific results, as illustrated by the early outcomes presented in Sections \ref{sec:overview_GP} and \ref{sec:CygX1}.

In addition to the standard GP scheduled observations, GP-ToO observations for astrophysical sources unrelated to GRBs are pre-planned each year. These ToO observations can be triggered when a cataloged source meets predefined criteria, such as exceeding a certain flux threshold, or being detected and/or monitored by other instruments. In addition, SVOM Co-Is and affiliated scientists can also request ToO observations on-the-fly when a particularly interesting astrophysical source becomes active or when a new source is discovered. This approach allows for timely monitoring of transient and variable phenomena while ensuring that observational resources are allocated efficiently. The ToO system is designed to perform observations within 48~h for standard ToOs and within 12~h for exceptional cases, such as Galactic supernovae or gravitational-wave alerts. In practice, these response times have been significantly reduced since the beginning of the SVOM operations, primarily due to the Beidou communication system, which allows ToOs to be uploaded to the satellite within minutes.
Standard ToOs (ToO-NOM) are typically observed for one orbit ($\sim$45~min), while exceptional ToOs (ToO-EX) can last up to 15 orbits. Multi-messenger ToO (ToO-MM), targeting counterparts of poorly localized events using optimized tiling for the narrow-field MXT and VT instruments, can perform up to five pointings per orbit, with a maximum of 75 tiles in total. 

During the nominal mission phase, 60\% of SVOM's effective observing time is dedicated to the GP, while 15\% is allocated to ToO observations. In the extended mission phase, the fraction of observing time devoted to ToO observations is expected to increase to 40\%, whereas the GP allocation will decrease to 35\%. In addition, the portion of GP observing time available outside the B1 law constraint is planned to increase substantially in the extended period of the mission, from 15\% to 50\%, allowing more time to be devoted to observations of the Galactic plane.

\subsection{Overview of Performed Observations During the First Year of Operations}\label{sec:overview_GP}

\begin{figure}
    \centering
     \includegraphics[width=1.0\linewidth]{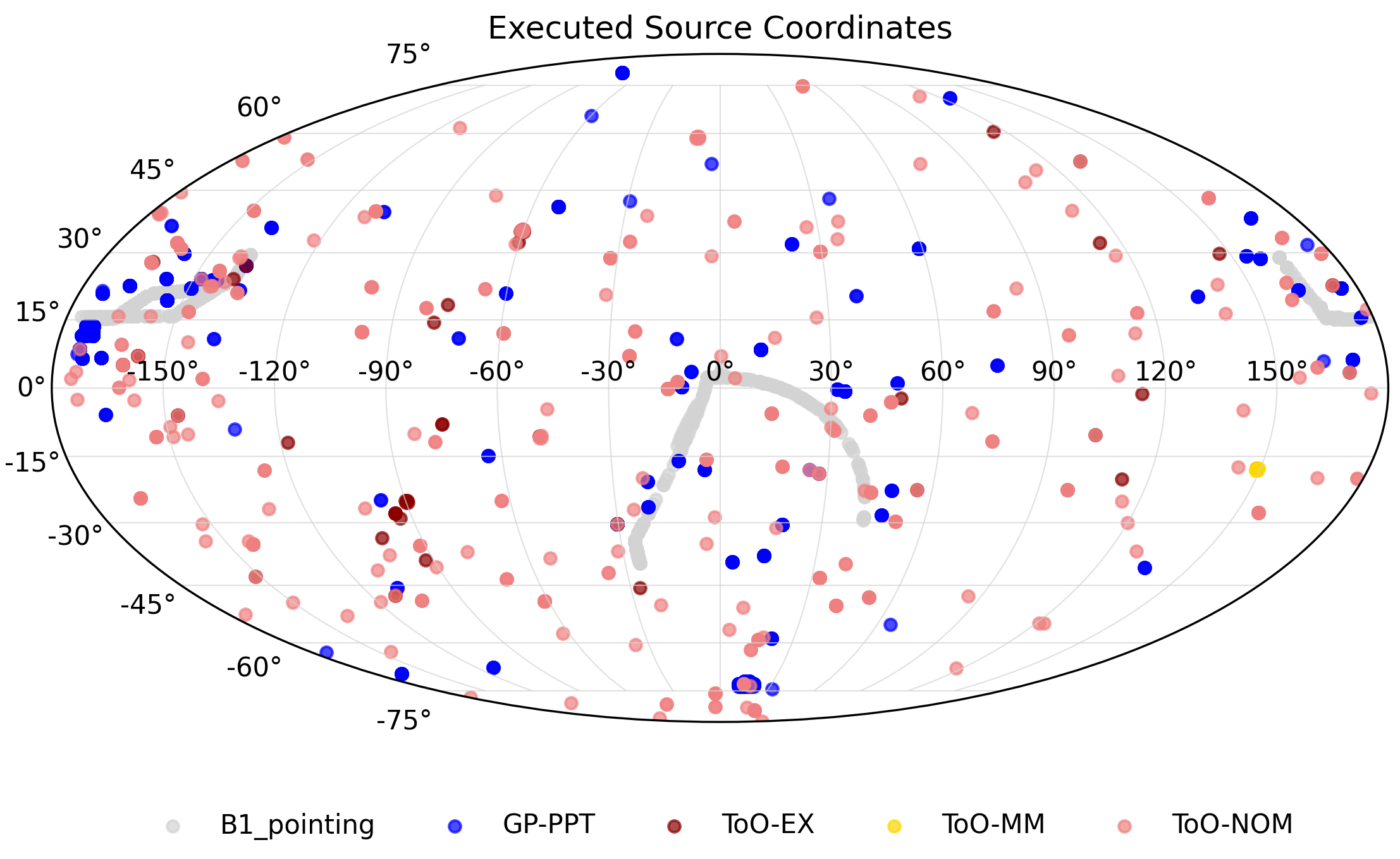}
    \caption{Distribution of the coordinates of observations of different GP and ToO types since the beginning of the operational phase of the mission (from Janury 15th 2025 to November 09th 2025).}
    \label{fig:GP_too_skymap_distribution}
\end{figure}

As of early November 2025, more than 1,200 GP observations and over 600 ToO observations had been scheduled. The sky distribution of these observations is shown in Figure~\ref{fig:GP_too_skymap_distribution}. Over this period, the total effective planned observation time amounted to approximately 10.2\,Msec \footnote{The effective planned observing time is defined as the total scheduled observing time multiplied by the mean MXT/VT observing availability per orbit, averaged over all declinations. This factor accounts for unavoidable time losses due to orbital constraints---such as Earth occultation, spacecraft slews, and other operational limitations---and therefore provides a realistic estimate of the exposure time available for scientific data acquisition, as opposed to the idealized total scheduled duration.}. The distribution of effective allocated time across the various observing programs is summarized in Table~\ref{tab:observation_summary}.

\begin{figure}
    \centering
     \includegraphics[width=1\linewidth]{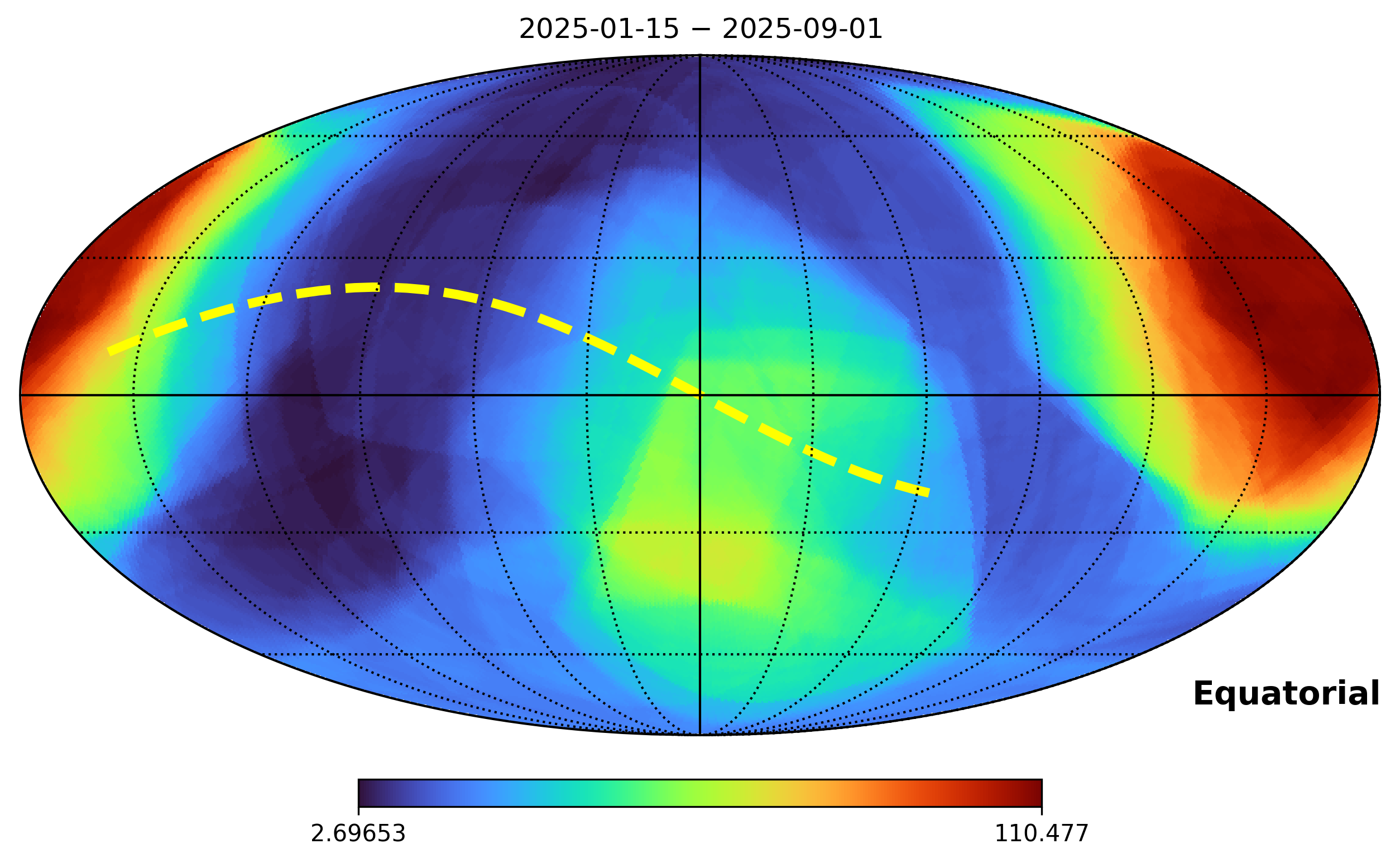}
    \caption{Effective ECLAIRs sky exposure map in days (from January 15th 2025 to September 1st 2025, in equatorial coordinates). The non-uniform efficiency across the ECLAIRs field of view is not taken into account. The yellow dashed line shows the Sun trajectory over the year.}
    \label{fig:GP_map_exp_com_valid}
\end{figure}

To assess how this planned time translated into actual scientific observations, one can examine the volume of data effectively obtained. For this purpose, we take the VT instrument as a reference case. During the same time interval, the VT successfully acquired approximately 7.6\,Msec of usable imaging data, corresponding to an overall effective completion rate of $\sim$74\%. The distribution of execution time and completion rates by source class is also presented in Table~\ref{tab:observation_summary}. Among GP observations, approximately 15\% of the allocated time was devoted to targets located more than $10^\circ$ from the satellite’s nominal B1 attitude law. Figure~\ref{fig:GP_map_exp_com_valid} shows the fields observed by ECLAIRs, illustrating that a large fraction of observing time was devoted to pointings aligned with the B1 attitude law, resulting in survey coverage concentrated near the Galactic poles.

In addition, between January and  November 2025, an average of 14.9 ToOs have been executed each week. Among them, 3.4 ToOs per week were performed on astrophysical targets related to the SVOM General Program and, more generally, on the Observatory Science for a total of 306 orbits (about 8 orbits per week).

\begin{table*}[h!]
\centering
\begin{tabular}{ccccccc}
\hline\hline
Category & Planning & Effective Planning & Execution & Effective Execution & Execution Proportion & Completion Rate \\
 & (nb. of targets) & (minute) & (nb. of targets) &  (minute) &  (\%) &  (\%)  \\
\hline
GP    & 1,288 & 56,691  & 850  & 43,957  & 34.9 & 77.5\\
ToO-EX    & 90 & 8,207  & 76   & 8,268   & 6.6 & 100.0 \\
ToO-NOM    & 544 & 34,268  & 456   & 32,574   & 25.9 & 95.1  \\
ToO-MM     &     & 61      &       &  46      & 0.04 & 74.9 \\
B1 pointing
& 574 & 71,315  & 426   & 41,104   & 32.6 & 57.6 \\
\hline
\end{tabular}
\caption{Summary of SVOM observing programs (2025 January 15–November 9). Listed are the number of sources, the effective planned and executed observing times, and the resulting completion fractions. B1 pointings are observations directed at sky positions compliant with the B1 attitude law, free of catalogued astrophysical sources within the narrow-FoV instruments’ field of view. \textit{Execution Proportion} indicates the percentage of total executed observation time accounted for by each observation category. \textit{Completion Rate} refers to the percentage of effectively executed observation time relative to the planned observation time for each category.}
\label{tab:observation_summary}
\end{table*}


\subsection{Example of first Scientific Results from the General Program: Long-Term Monitoring of the Microquasar Cygnus X-1}\label{sec:CygX1}

Cygnus X-1 (a.k.a. Cyg X-1) is a high-mass X-ray binary located at a distance of $\sim$2\,kpc, consisting of the O9.7 Iab supergiant HD 226868 orbiting a $\sim$10 M$_\odot$ compact object \citep{Bolton72, Stark77}, and is widely regarded as the prototype stellar-mass black-hole system. Cyg  X-1 was observed through dedicated pointings during the commissioning and verification phases, as well as through a dedicated monitoring campaign conducted within the framework of the GP in 2025. A total of 47 pointed observations\footnote{An observation is defined here as a continuous time interval during which the platform attitude, and thus the pointing coordinates, remain fixed.} were carried out, resulting in a cumulative exposure of about 250 ks.

    
We extracted spectra for each pointed observation in order to study the variability of the source in terms of flux and spectral shape, with the aim of investigating possible state changes between July 17th 2024 and July 21th 2025. Spectral fits performed with phenomenological models in \textsc{xspec}) do not reveal any significant variations in the spectral parameters, which is consistent with the light curves provided by all-sky monitors such as the BAT instrument onboard the Neil Gehrels Swift Observatory (hereafter Swift; \citealt{2004ApJ...611.1005G}) \footnote{\url{https://swift.gsfc.nasa.gov/results/transients/CygX-1/}} and MAXI/GSC \footnote{\url{https://maxi.riken.jp/star_data/J1958+352/J1958+352.html}}\citep{2011PASJ...63S.623M}.

\begin{figure}
        \centering
        \includegraphics[width=\linewidth]{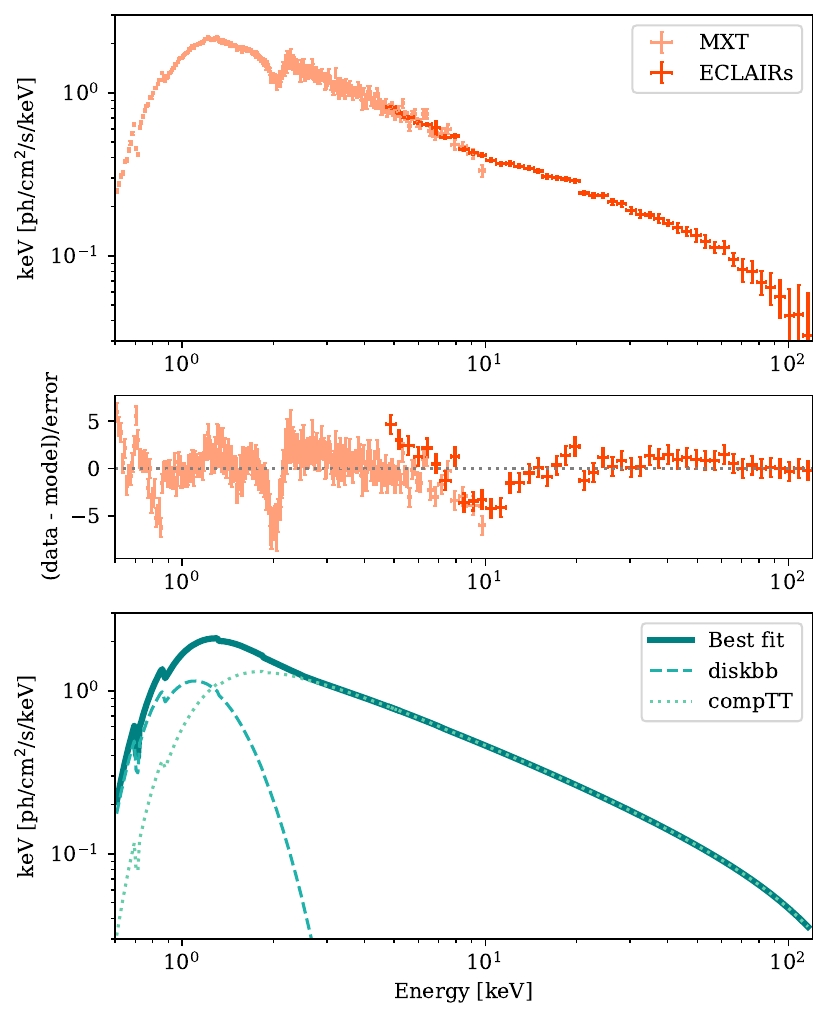}
        \caption{Joint MXT and ECLAIRs spectral fit of Cygnus X-1. The data are fitted with the model \textsc{constant*tbabs(diskbb+comptt)}. Top: MXT (coral) and ECLAIRs (orange) spectra in units of keV~[ph~cm$^{-2}$~s$^{-1}$~keV$^{-1}$]. Middle: fit residuals. Bottom: best-fit model (solid line), with the \textsc{diskbb} and \textsc{comptt} components shown as dashed and dotted lines, respectively.}
        \label{fig:spec_fit_cygx-1_1}
    \end{figure}

Since the source does not show significant variability during the observation period, we stacked the spectra from these observations with MXT and ECLAIRs to perform a joint fit with both instruments over the 0.6--120\,keV energy band. For this purpose, we adopted in \textsc{xspec} the model \textsc{constant*tbabs(diskbb+comptt)}, which includes both an accretion disk component \citep[\textsc{diskbb},][]{Mitsuda1984} and a Comptonization component \citep[\textsc{compTT}][]{Titarchuk1994} produced by the interaction of seed photons originating from the disk, with a hot electron plasma located close to the black hole. Therefore, the temperature at the inner disk radius ($kT_{in}$) is linked to the seed photon temperature ($kT_{0}$). 
The \textsc{tbabs} component accounts for Galactic absorption, while the \textsc{constant} factor accounts for cross-calibration between the two instruments. The calibration constant is fixed at 1 for MXT, while the ECLAIRs constant is left free to vary. The spectra, the best model obtained from the fit, and the residuals are shown in Fig. \ref{fig:spec_fit_cygx-1_1}.

The derived spectral parameters are consistent with the typical hard state of Cygnus X-1, in agreement with previous studies reported in the literature \citep[e.g.,][]{Rodriguez2015, Cangemi2021}. We measure an inner disk temperature of $kT_\mathrm{in} = 0.252 \pm 0.003$\,keV, along with a coronal optical depth $\tau = 1.59^{+0.24}_{-0.43}$ and an electron temperature of $29^{+11}_{-4}$\,keV. The cross-calibration constant for ECLAIRs is found to be $0.89 \pm 0.02$. The residuals observed below 2\,keV are most likely attributable to limitations in the current response matrix file (RMF), which is still undergoing refinement.

\section{Serendipitous High-Energy Source Detection and Monitoring with SVOM/ECLAIRs} \label{sect:SHES}

The large FoV and continuous monitoring of the high-energy sky provided by ECLAIRs give SVOM a unique capability to detect transient and variable sources beyond its core program dedicated to GRB detection. Throughout the first year of operations, ECLAIRs routinely recorded serendipitous high-energy activity from sources located within its FoV, yielding a rich dataset of bursts, flares, and enhanced emission episodes across multiple source classes. These detections rely on a combination of onboard triggering, ground-based reprocessing, and systematic quick-look analysis of the X-band data, enabling the identification of both known and previously unreported transients. Together, these tools form a coherent framework for time-domain high-energy astronomy. In this section, we describe the trigger and analysis architecture supporting serendipitous detections and present the first scientific results obtained from non-GRB events discovered by ECLAIRs.

\subsection{Triggering and Analysis Framework for Serendipitous High-Energy Source Detections}
\label{sect:triggers}

Serendipitous high-energy transients detected by ECLAIRs can be identified through three complementary approaches. Rapid, short-timescale events are detected either onboard by the \textit{Unité de Gestion et de Traitement Scientifique} (UGTS, Scientific Trigger and Control Unit, \citealt{Schanne+etal+2026}, see Section \ref{sec:ugts}) trigger system or through ground-based reprocessing using the Offline Trigger (\citealt{Brunet+etal+2026}, see Section \ref{sec:oft}), while longer-timescale transients and variable sources are primarily identified and monitored through the systematic Quick-Look Analysis (QLA) based on the analysis of the ECLAIRs X-band data (\citealt{Goldwurm+etal+2026}, see Section \ref{sect:EQLA}) . Once a transient or significant high-energy activity unrelated to GRB events is detected by any of these channels, the information is handled within the dedicated Serendipitous High-Energy Sources (SHES) working group. The SHES scientists on duty are responsible for assessing the event, disseminating alerts via ATels and/or GCN notices when appropriate, and coordinating follow-up observations with SVOM or other facilities through ToO requests.

\subsubsection{The UGTS Onboard Trigger System}
\label{sec:ugts}

The onboard processing of ECLAIRs data, including the trigger algorithms responsible for detecting and localising both GRBs and known sources, is performed by the \textit{Unité de Gestion et de Traitement Scientifique} (UGTS, Scientific Trigger and Control Unit, \citep{Schanne+etal+2026}. To detect transient events, the UGTS operates two complementary trigger algorithms in parallel:
\begin{itemize}
    \item The \textit{count-rate trigger} (CRT) monitors variations in the detector count rates on timescales ranging from 10\,ms to 20.48\,s. When a significant excess is detected, a sky image is reconstructed over the corresponding interval to identify the source location. This method is particularly effective for identifying rapid, short-lived bursts.
    
    \item The \textit{image trigger} (IMT) continuously reconstructs sky images from 20.48 s exposures, which can be stacked over progressively longer timescales. This approach extends the sensitivity to transients lasting up to about 20 minutes, thereby complementing the CRT by enhancing the detection of slower phenomena.
\end{itemize}

The UGTS use a catalogue of known X-ray sources \citep{DagoneauCatalogue}, which serves several purposes:
\begin{itemize}
    \item enabling the subtraction of the brightest sources (mostly in the galactic plane, plus Sco X-1) to reduce the coding noise they introduce;
    \item guiding the avoidance strategies applied during GRB searches;  
    \item enabling the detection of outbursts when catalogued (but unsubtracted) sources undergo flaring activity.  
\end{itemize}

The way activity from known sources is handled depends on their catalogue status. When the IMT detects activity from catalogued but unsubtracted sources, the system generates a \textit{CAT} alert sequence. This mechanism proved effective very early in the mission, with the first alert issued for the pulsar XTE J1946+274 in July 2024 \citep{2024ATel16719....1R}. Conversely, if either CRT or IMT identifies activity from a source not listed in the catalogue, a \textit{GRB} alert sequence is issued. In the latter case, the association with a known source, if any, is established during ground-based analysis by the Burst Advocate Team  \citep{Cordier+etal+2026, Claret+etal+2026} and further analysis and monitoring is then performed by the SHES scientists on duty.

The UGTS is highly configurable, with numerous adjustable parameters controlling its behaviour. One such option is the automatic satellite slew in response to CAT alerts, which can be enabled or disabled. As of September 2025, this functionality is disabled  to restrict the slew to GRB alert sequences exclusively . In addition, each source in the onboard catalogue is assigned a dedicated detection threshold for triggering CAT alerts. These thresholds are adaptive: they are raised automatically onboard to prevent repeated triggers on similar excesses and reset either when a new configuration is uploaded or following an ECLAIRs UGTS reboot.

\subsubsection{The Offline Trigger (OFTG)}

\label{sec:oft}

In addition to the onboard UGTS, ECLAIRs data are reprocessed at the ECLAIRs instrument center using the Offline Trigger (OFTG; \citealt{Brunet+etal+2026}). Working on ground on the X-band data composed of all photons recorded onboard, it benefits from higher computational resources than the onboard trigger and is thus designed to detect a wide range of transient and variable phenomena. The OFTG can be operated in three distinct modes, depending on the class of transient being targeted:

\begin{itemize}

    \item \textit{Blind search.} This mode is primarily used to detect GRBs, thermonuclear type~I X-ray bursts and stellar flares. It combines three complementary trigger algorithms: (i) a count-rate–based trigger, implementing three background subtraction techniques—a simple polynomial fit (as done onboard), a Fast Fourier Transform (FFT) method, and a wavelet-based method; (ii) an image-based trigger, operating on exposure times starting at 2.56\,s; and (iii) a machine-learning–based trigger relying on anomaly-detection algorithms, which is currently under development.

    \item \textit{Targeted search.} This mode is used to assess whether transients reported by external facilities were detected by ECLAIRs. It employs the same trigger algorithms as the blind search, but is restricted to predefined time intervals and sky regions.

    \item \textit{Millisecond search.} This mode is dedicated to the detection of msec-timescale duration events, such as terrestrial gamma-ray flashes (TGFs) and magnetar flares, including non-localizable events. It relies exclusively on count-rate–based detection methods. Three techniques have been implemented to date: Poisson probability analysis, Bayesian blocks, and a wavelet-based method.

\end{itemize}

As in the UGTS, the OFTG makes use of a catalog of known sources to mitigate coding noise and to facilitate the identification of source-related activity.
The OFTG is highly configurable, allowing its parameters to be tailored to the transient class of interest, in particular with respect to timescales, energy bands, and source subtraction. This flexibility enables the detection of both short- and long-duration transients, as well as the systematic monitoring of individual sources. Unlike the UGTS, the OFTG does not apply source-specific detection thresholds; instead, a single threshold is defined for each timescale in order to identify significant activity from any catalogued source.

\subsubsection{Quick-Look Analysis (ECPI/QLA)}
\label{sect:EQLA}

As described in \citep{Goldwurm+etal+2026}, the ECLAIRs data reduction and analysis pipeline (ECPI) is run automatically in Quick-Look Analysis (QLA) mode upon reception of the X-band data on the ground. It generates high-level scientific products for each observation period, with a latency primarily resulting from the delay to downlink the data from the plateform through X-Band antennas. 
The pipeline runs on the SVOM French Science Center (FSC) infrastructure \citep{Louvin+etal+2026} using a default configuration with five energy bands: 4–10, 10–20, 20–40, 40–80, and 80–150\,keV.

Moreover, since no trigger is used for this analysis, the QLA integrates all data from the good-time intervals of each observation. Its primary goal is to monitor sources displaying outbursts on timescales longer than those captured by the UGTS and OFTG triggers, as well as to track spectral variability.

A dedicated web interface provides access to QLA products for the collaboration, generating figures that include both low-level information—such as count-rate evolution and auxiliary data—and high-level results, including sky maps and source spectra. This interface allows the quality of the data reduction to be assessed at each step, helping to rule out false detections and enabling rapid identification of source outbursts. In addition to catalogued sources, ECPI can detect previously unknown sources appearing in the FoV; while no such events had been recorded at the time of writing, the community could be quickly alerted of the appearance of a new transient source.

Moreover, count-rate light curves are produced with time bins of 1.5\,h (1 orbit), 6\,h, and 1\,day across all five energy bands. Although the current SVOM observing strategy does not include a daily near–full-sky survey comparable to the Swift/BAT transient monitor \citep{Krimm_2013}, any pointing covering the Galactic plane allows flux measurements of all bright Galactic sources within the FoV.

\subsection{First Serendipitous Detections and Monitoring Results}

\begin{figure*}
        \centering
        \includegraphics[width=1.0\linewidth]{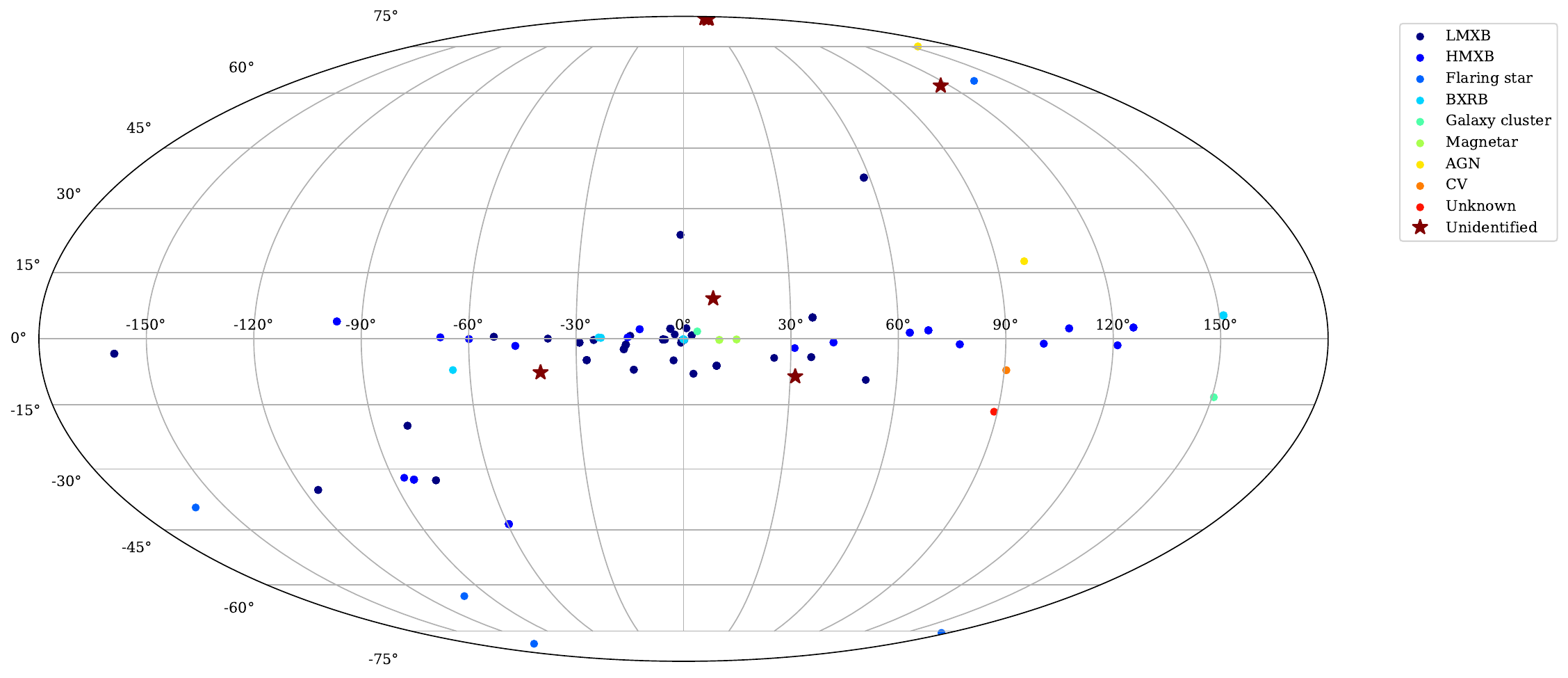}
        \caption{Sky map of non-GRB triggers detected by ECLAIRs between June 22nd 2024 and August 31st 2025. Colors indicate different source classes, and brown stars denote triggers of unidentified origin. One source is marked as “unknown”, as it originates from the ROSAT catalogue but lacks a proper astrophysical classification.}
        \label{fig:map_non_grbs_triggers}
    \end{figure*}
\begin{figure}
        \centering
        \includegraphics[width=1\linewidth]{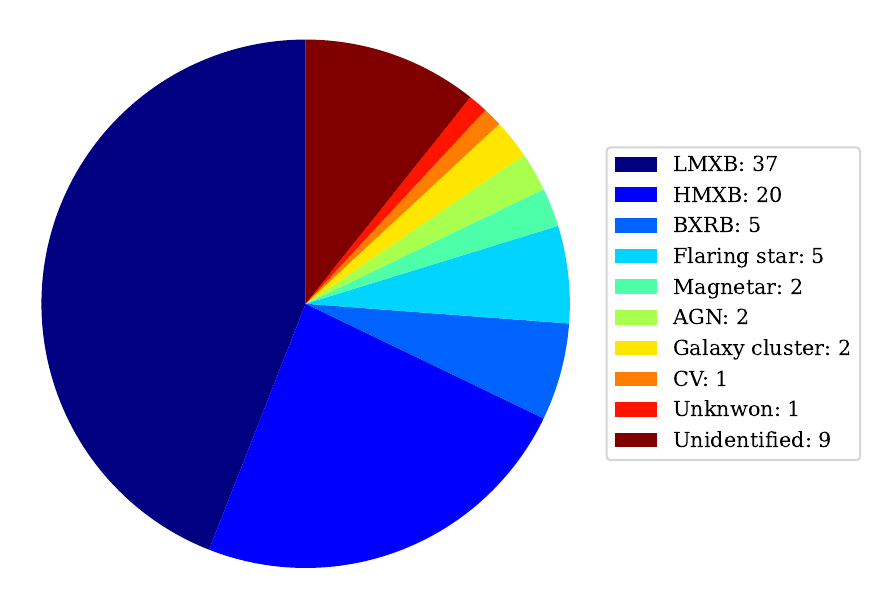}
        \caption{Pie chart of non-GRB sources detected by ECLAIRs, classified by source type.}
        \label{fig:pie}
    \end{figure}

Figure \ref{fig:map_non_grbs_triggers} shows the sky distribution of non-GRB triggers detected between June 22nd 2024 and August 31st 2025, color-coded by source class, while Figure \ref{fig:pie} shows the fraction of triggers by type. For a small subset of events, it remains unclear whether the observed excess corresponds to a genuine increase in source flux or to a background fluctuation.

The vast majority of non-GRB triggers (273 in total) are associated with X-ray binaries. Approximately two-thirds of these events originate from Low-Mass X-ray Binaries (LMXBs), yielding 174 triggers, most of which correspond to type~I thermonuclear X-ray bursts (examples of such detections are given in Sections \ref{sec:4U} and \ref{sec:Aql}). High-Mass X-ray Binaries (HMXBs) account for 62 additional triggers, while 20 events are associated with candidate Black Hole X-ray Binaries (BXRBs). In this paper, the terms LMXB and HMXB are used specifically to refer to neutron star systems. In addition, other triggers are related to AGN and stellar flares (as reported in Sections \ref{sec:1ES} and \ref{sec:RSCVn}). Finally, nine triggers are linked to unidentified sources, five of which have positions compatible with the Coma cluster. These events, described in Sections~\ref{sec:Coma} and \ref{sec:unidentified}, are indicated by orange stars in Figure~\ref{fig:map_non_grbs_triggers}, and their positions are listed in Table~\ref{tab:unidentifed_trigger}.

\begin{table*}[h!]
\centering
\begin{tabular}{cccccc}
\hline\hline
Burst\_id & Trigger time & RA (deg) & DEC (deg) & SNR & Localisation error (arcmin) \\
\hline
sb25020104    & 2025-02-01 14:08:20.314000 & 235.416  & -64.743  & 10.84  & 14.36 \\
sb25021804    & 2025-02-18 07:20:27.858000 & 195.021  & 28.078   & 7.56   & 19.43 \\
sb25030301    & 2025-03-03 01:50:29.001000 & 194.595  & 27.871   & 8.20   & 18.13 \\
sb25032302    & 2025-03-23 04:08:21.624000 & 194.837  & 28.038   & 7.90   & 18.72 \\
sb25042207    & 2025-04-22 22:36:17.357000 & 194.927  & 27.774   & 7.01   & 20.74 \\
sb25050907    & 2025-05-09 20:29:25.993000 & 195.134  & 27.979   & 7.30   & 20.03 \\
sb25061218    & 2025-06-12 20:36:00.000000 & 198.503  & 54.658   & 7.01   & 20.74 \\
sb25062615    & 2025-06-25 04:16:53.871000 & 289.785  & -5.221   & 6.60   & 21.87 \\
ECL\_250628\_1111 & 2025-06-28 11:11:26.600000 & 262.708  & -17.047  & 15.15  & 11.05 \\
\hline
\end{tabular}
\caption{List of unidentified non-GRB triggers detected by ECLAIRs.}
\label{tab:unidentifed_trigger}
\end{table*}


\subsubsection{Burst Oscillations in a Type I X-ray Burst of LMXB 4U 0614+091 and Monitoring of a New X-ray Outburst}\label{sec:4U}

The ECLAIRs onboard trigger system issued an alert on January 10th 2025 at 15:58:03 UTC, consistent with the coordinates of the LMXB 4U\,0614+091, an ultra-compact binary consisting of a neutron star and a white dwarf. The spectral analysis of this event using ECPI \citep{Goldwurm+etal+2026} identifies it as a classical Type I thermonuclear burst, a phenomenon previously observed from this source (see e.g. \citealt{bat_discovery}).

\begin{figure}
        \centering
        \includegraphics[width=1\linewidth]{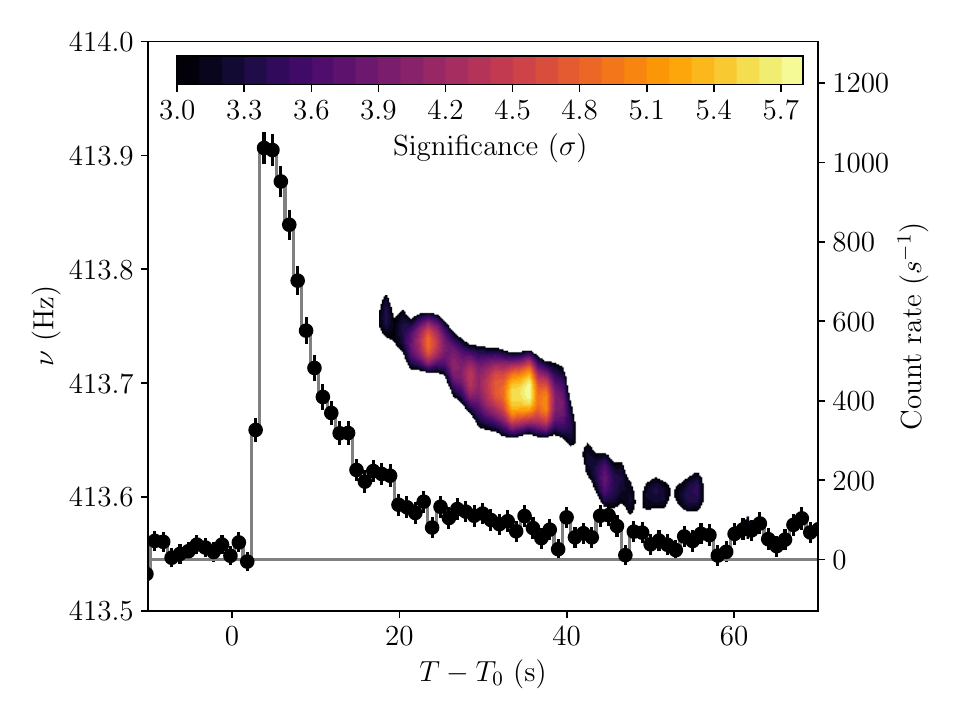}
        \caption{Significance contours of oscillation detections during the Type I X-ray burst of 4U~0614+091 are shown as a function of time and frequency (left-hand axis). These were computed using a $Z_1^2$ test applied in a 20-second sliding window with 1-second steps. The significance levels are represented by a color gradient. The burst light curve, shown on the right-hand axis as the equivalent on-axis count rate in the 4–40 keV energy band, has the persistent emission subtracted and is binned in 1-second intervals. Adapted from \citet{LeStum+etal+2026}.}
        \label{fig:osc_4U}
    \end{figure}

The time-integrated spectrum over the burst duration is well described by a blackbody model (\textsc{bbodyrad}), with a temperature of $kT = 2.04 \pm 0.04$~keV and an apparent emission radius of $R = 5.8 \pm 0.3$~km, assuming a source distance of 3\,kpc. This distance corresponds to an average of the value estimated by \citet{Galloway_2020_minbar} and that derived from the GAIA parallax \citet{2021MNRAS.502.5455A}. The burst reaches a peak flux of $1.39 \pm 0.04 \times 10^{-7}$\,erg\,cm$^{-2}$\,s$^{-1}$ and has a time-integrated bolometric energy of $(1.89 \pm 0.04) \times 10^{39}$ erg, values typical of Type I bursts.
We conducted a search for burst oscillations, motivated by their previous detection with Swift/BAT \citep{bat_4U0614} at 414.75 Hz in a 2006 burst and with GECAM at 413.63 Hz \citep{gecam_4U0614}. A significant oscillation is detected between 11\,s and 62\,s after the trigger, exhibiting a systematic downward frequency drift with $\dot{\nu} = (-4.7 \pm 0.3) \times 10^{-3}$~Hz\,s$^{-1}$ (see Figure \ref{fig:osc_4U}). The average frequency is $\bar{\nu} = 413.674 \pm 0.002$~Hz, which differs from the value reported by Swift/BAT. We tentatively attribute both the frequency offset and drift to Doppler modulation caused by the neutron star orbital motion. Under this interpretation, the orbital period of the system cannot exceed 20 minutes, irrespective of other orbital parameters.
These results demonstrate that ECLAIRs is well suited for the study of Type I bursts from LMXBs, highlighting the potential of the SVOM Observatory Science program to contribute significantly to this field. A more detailed analysis is presented in \citet{LeStum+etal+2026}.

\begin{figure*}
        \centering
        \includegraphics[width=1\linewidth]{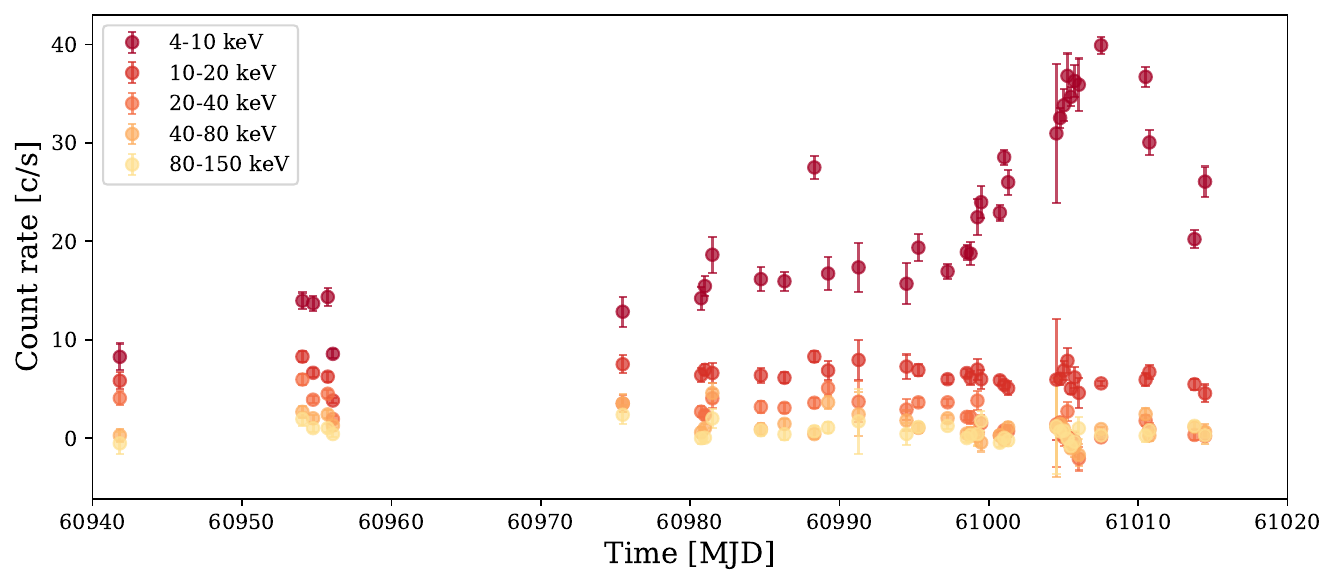}
        \caption{ECLAIRs hard X-ray lightcurve of the November 2025 outburst of 4U~0614+091 as provided by the ECPI/QLA in five different energy bands.}
        \label{fig:4U_LC}
    \end{figure*}

Following the January 2025 outburst, renewed X-ray activity from 4U~0614+09 was detected with the ECLAIRs QLA web interface starting in early November 2025 \citep{2025ATel17528....1R}. The source flux increased substantially, rising from $\sim$14 counts~s$^{-1}$ to $\sim$40 counts~s$^{-1}$ in the 4–10~keV energy band by 28 November 2025 (see Fig.~\ref{fig:4U_LC}). A preliminary spectral analysis was performed on the December 1st 2025 observation, during which the source was serendipitously in the ECLAIRs FoV. The light curve showed a declining trend from $36.7 \pm 0.6$ to $30.0 \pm 0.7$ counts~s$^{-1}$ over the course of a day, and the spectrum was extracted from a 15.7~ks exposure.
The resulting spectrum is well described ($\chi^2 = 17.07$ for 19~d.o.f.) by a model comprising an unabsorbed blackbody component with temperature $kT = 1.3 \pm 0.1$~keV and radius $R~=~1.92 \pm 0.56$~km (assuming a distance of 3~kpc), and an additive power-law component characterized by photon index $\Gamma = 2.9^{+0.2}_{-0.4}$ and normalization $2.7^{+3.0}_{-1.9}$. The derived unabsorbed flux in the 4–50~keV band is $2.17^{+0.09}_{-0.07} \times 10^{-9}$~erg~cm$^{-2}$~s$^{-1}$.
No type I thermonuclear bursts were observed during the rising phase, consistent with the source’s known behavior of outbursts with highly variable recurrence times and occasional burst oscillations.

    \subsubsection{Spectral-State Monitoring of Aql X-1 During its 2024 Outburst}\label{sec:Aql}

\begin{figure}[h!]
        \centering
        \includegraphics[width=\linewidth]{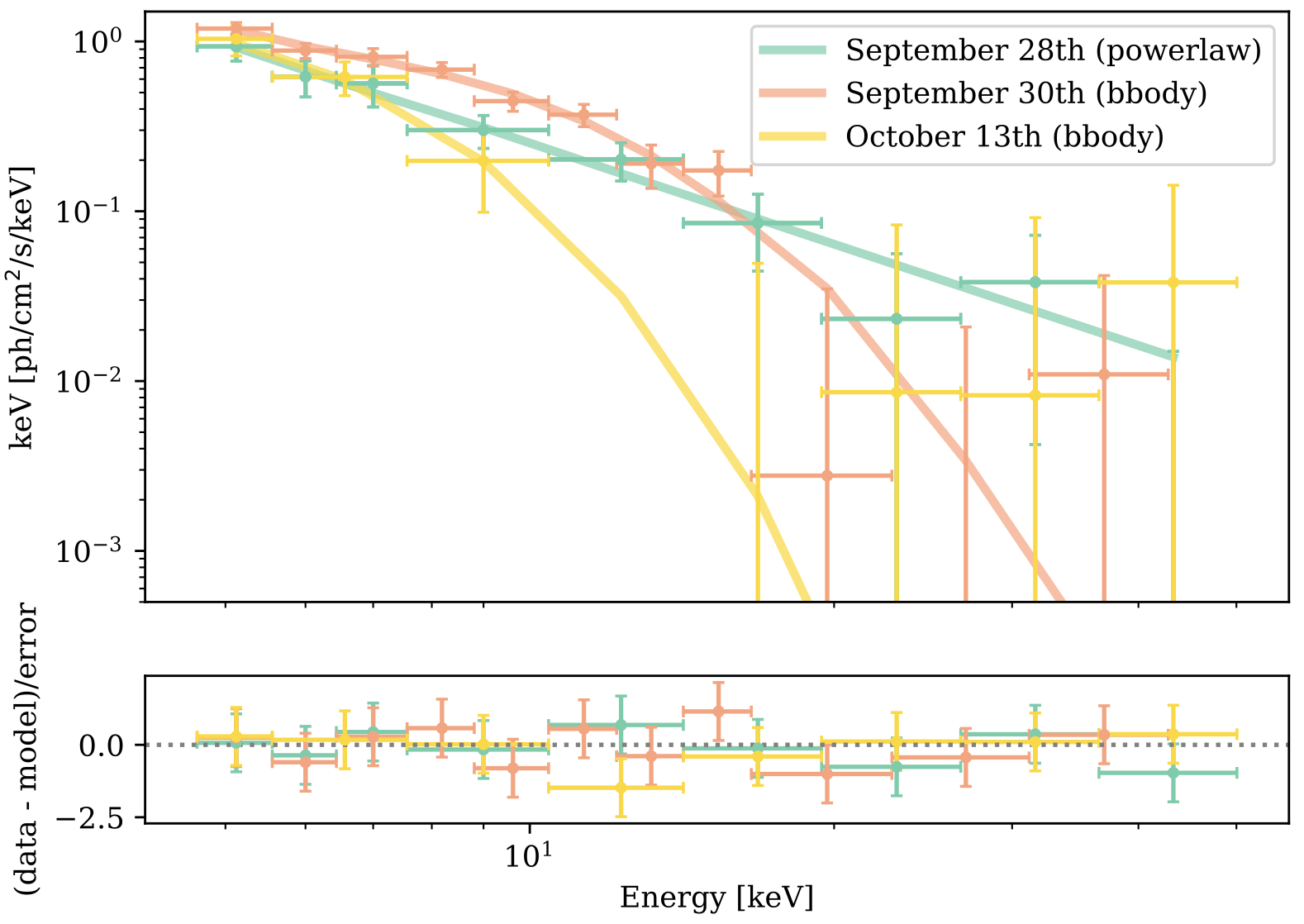}
        \caption{ECLAIRs spectra of Aql~X-1 at three epochs during the spectral state transition monitored by SVOM/ECLAIRs in september-october 2024.}
        \label{fig:aqlX1}
    \end{figure}

The SVOM/ECLAIRs monitoring of the high-energy sky provided an early and sensitive view of the recent outburst of the neutron-star low-mass X-ray binary Aql X-1 which occured in September--October 2024. The source was clearly detected in the ECLAIRs QLA data (see section \ref{sect:EQLA}) at the onset of the event. This enabled SVOM to identify the rise of the outburst nearly simultaneously with the earliest alerts reported by other facilities, including Einstein Probe \citep{2024ATel16821....1L}. The initial ECLAIRs detections showed a hard spectral shape, consistent with the canonical hard state typically observed during the early rising phase of Aql X-1 outbursts (see Figure \ref{fig:aqlX1}).

As the outburst evolved, the series of ECLAIRs observations revealed a gradual softening of the high-energy emission (see Figure \ref{fig:aqlX1} and \citealt{2024ATel16843....1L}). The inferred spectral evolution shows that the source transitioned from a hard, power-law–dominated spectrum toward a softer configuration, in agreement with state changes reported in contemporaneous observations by Einstein Probe, NICER, and NuSTAR \citep{2025arXiv251116437M}. The observed change in hardness ratio, combined with the temporal evolution of the flux, is fully consistent with the well-established hard-to-soft transition that characterizes the accretion-state evolution of Aql X-1.

ECLAIRs continued to detect the source throughout the decay phase of the event. Approximately one month after the soft-state epoch, the high-energy emission hardened again and the flux dropped to a low level, indicating a return to the quiescent (hard) state. This re-hardening, together with the significant decrease in count rate, marks the completion of a full outburst cycle. These results demonstrate that ECLAIRs, despite its primary role as a trigger instrument, is capable of providing scientifically valuable monitoring of transient neutron-star systems, offering coverage that is complementary to the observations from Einstein Probe, NICER, and NuSTAR.

\subsubsection{X-ray Outburst of the Blazar 1ES 1959+650}\label{sec:1ES}
On December 6th 2024, SVOM/ECLAIRs observed a strong X-ray outburst from 1ES 1959+650, a high-synchrotron-peaked BL Lac object and one of the brightest X-ray blazars known, marking SVOM’s first detection of an AGN flare \citep{Coleiro_ATEL_1ES}. The rapid onboard trigger enabled SVOM ToO observations less than two hours later with MXT and VT, revealing the source in a significantly enhanced state across both soft X-ray and optical bands. A coordinated monitoring campaign followed: SVOM observed the event through December 26th 2024, while Swift initiated complementary observations on  December 12th 2024, continuing until March 1st 2025. Together, these two missions provided dense temporal coverage over nearly three months, spanning from optical/UV wavelengths up to 50 keV as reported in  \citet{2025ATel16978....1F, 2024ATel16941....1K, 2024ATel16955....1K, Foisseau+etal+2026}.  

\begin{figure}
        \centering
        \includegraphics[width=1\linewidth]{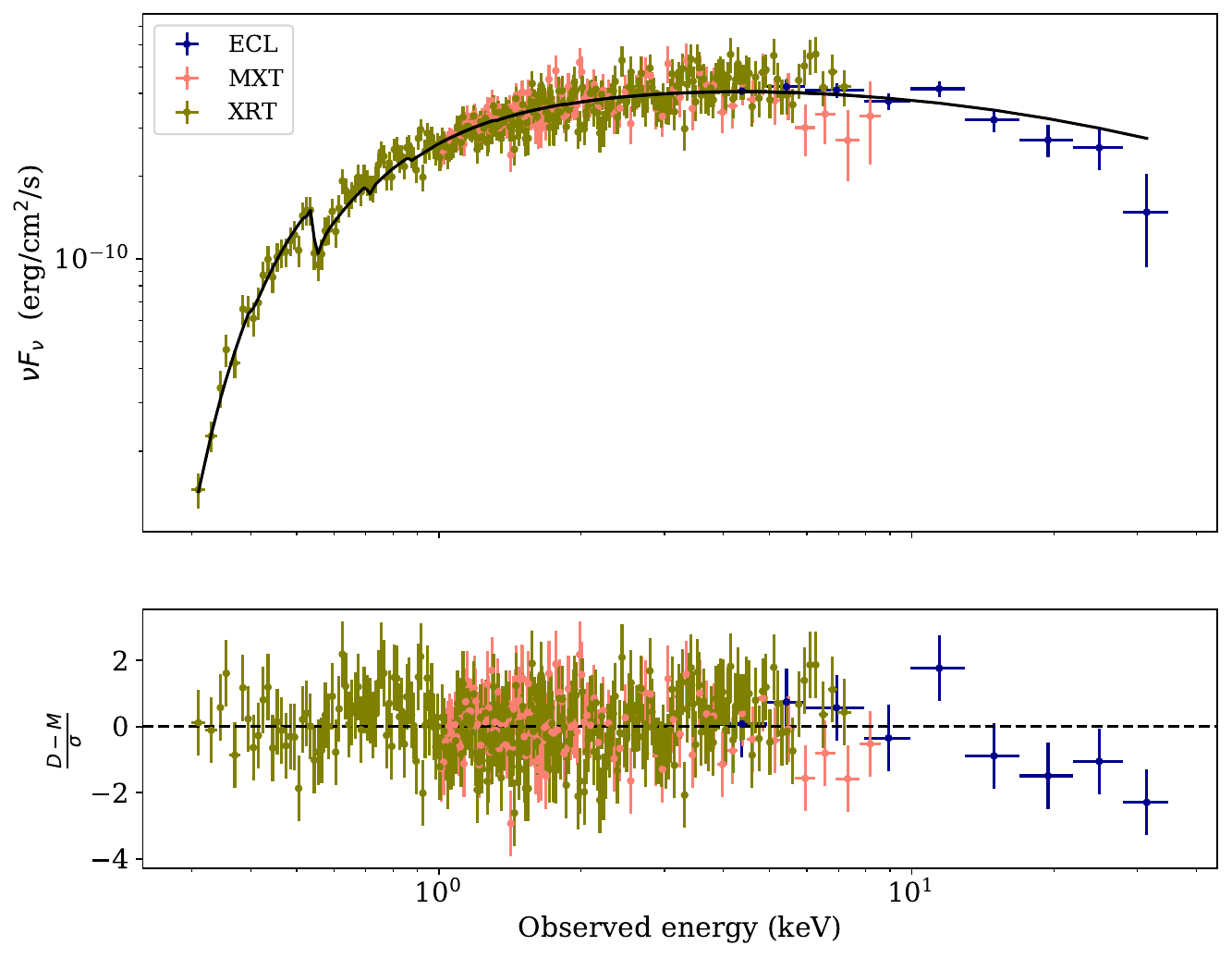}
        \caption{1ES 1959+650 combined Swift/XRT, SVOM/MXT and ECLAIRs spectra obtained on December 16, 2024. The bottom panel shows the residuals, computed as the difference between the observed data ($D$) and the model ($M$), normalized by the error ($\sigma$). Adapted from \citet{Foisseau+etal+2026}.}
        \label{fig:1ES_spectra}
    \end{figure}

The spectral analysis of MXT, ECLAIRs, and Swift/XRT data (see Fig.~\ref{fig:1ES_spectra}),  with an absorbed log-parabola model\footnote{The log-parabola spectral model was defined as $F(E)~=~K~\left( \frac{E}{E_1} \right)^{-\left[\alpha + \beta \log(E/E_1)\right]}$~ph\,cm$^{-2}$\,s$^{-1}$\,keV$^{-1}$ where $E_1$ is the pivot energy fixed 1~keV, $\alpha$ is the photon index at 1~keV , $\beta$ is the  curvature parameter, and  $K$ the  flux normalization.} with the absorption column density fixed at the Galactic value, shows that the flare exhibited strong X-ray variability with a pronounced harder-when-brighter trend. The synchrotron peak energy shifted to higher energies at higher flux levels, consistent with enhanced particle acceleration during the brightest phases. 

Correlation and spectral-evolution diagnostics—including $\alpha$–flux, $\beta$–flux, $E_{\rm p}$–$\beta$, $E_{\rm p}$–$S_{\rm p}$, and hardness-ratio loops—revealed no clear dominance of either Fermi-I shock acceleration or Fermi-II stochastic acceleration during the flare. While the positive correlation between synchrotron peak energy and peak flux is compatible with stochastic acceleration, the absence of a clear $E_{\rm p}$–$\beta$ anti-correlation and the lack of significant clockwise or counter-clockwise loops in the hardness-ratio/soft X-ray flux plane argue against a single dominant process. These results suggest a mixed Fermi-I/II scenario, likely reflecting complex and evolving physical conditions within the jet \citep{Foisseau+etal+2026}.  

This study demonstrates SVOM’s capability to detect bright X-ray blazar outbursts in real time and to trigger rapid multi-instrument follow-up. The combined SVOM–Swift campaign underscores the missions’ strong complementarity, particularly SVOM/ECLAIRs’ coverage below 10 keV, which extends spectral constraints beyond what Swift/BAT alone can provide. This first blazar outburst detected by SVOM paves the way for deeper investigations of particle acceleration in blazar jets.

\subsubsection{Chromospheric Evaporation in the RS CVn Star HD 22468}\label{sec:RSCVn}

A hard X-ray transient was detected by SVOM/ECLAIRs on January 9th 2025 at 11:39:01.2 UTC. The transient had a signal-to-noise ratio of 9.47 and was localized at R.A. = $\mathrm{03^h36^m39^s}$ and Dec. = $+00\degr33\arcmin25\arcsec$, with a positional uncertainty of $8\farcm4$. Within this error region lays the bright X-ray star HD~22468 (1RXS~J033647.2+003518; 3XMM~J033647.2+003515), an RS CVn–type binary system located approximately $4\arcmin$ from the ECLAIRs best-fit position. Even though hard X-ray emission from active stars is less commonly detected than soft X-ray emission ($<$10 keV), such high-energy ($>$10 keV) radiation has previously been reported during stellar flares (see e.g. \citealt{Osten2007, Osten2010, Tsuboi2016, Hakamata2025, Urabe2026}).
The identification of the transient as a stellar flare from HD~22468 was supported by contemporaneous multiwavelength observations. The field was simultaneously monitored by the SVOM/GWAC cameras, which detected a white-light (WL) flare from HD~22468 peaking approximately one hour after the ECLAIRs trigger. In addition, follow-up spectroscopy obtained with the National Astronomical Observatory of China (NAOC) 2.16-m telescope at $\sim$1.7\,hr after the trigger revealed enhanced H$\alpha$ emission originating from the upper chromosphere. Owing to the limited spectral resolution ($R\sim2900$) and the lack of coverage of the Ca\,{\sc ii} H and K lines, the widely used $S$-index cannot be measured. Instead, we adopt the secondary H$\alpha$ activity index, $R_{\mathrm{H\alpha}}$, as a proxy for stellar activity (see a summary of $S$-index measurements from multiple emission lines in \citealt{Mignon2023}).

The $R_{\mathrm{H\alpha}}$ index is defined as the ratio of the stellar H$\alpha$ line luminosity to the bolometric luminosity,
\begin{equation}
R_{\mathrm{H\alpha}}=\frac{L_{\mathrm{H\alpha}}}{L_{\mathrm{bol}}}
=\left(\frac{f_{\mathrm{H\alpha}}}{\sigma T^{4}}\right)
\left(\frac{d}{R_{\star}}\right)^{2},
\end{equation}
where $f_{\mathrm{H\alpha}}$ is the H$\alpha$ line flux, $T$ is the stellar effective temperature, and $d$ and $R_{\star}$ are the stellar distance and radius, respectively (see e.g., \citealt{He2023, Li2023, Li2024L}). Using the parameters reported in \citet{Wang+etal+2026}, we obtain $R_{\mathrm{H\alpha}} = 9.1\times10^{-5}$, which is at least an order of magnitude higher than typical values for inactive main-sequence stars and giants \citep{He2023, He2025}.

Spectral modeling of the hard X-ray emission with an \textsc{apec} thermal plasma model indicated a peak temperature of $106^{+27}_{-22}$~MK and a total bolometric energy release of $\sim1.7\times10^{38}$~erg, comparable to that inferred from the associated WL flare. Furthermore, the H$\alpha$ emission line exhibited a bulk blueshift of $-96\pm20\ \mathrm{km\ s^{-1}}$ relative to the stellar photosphere during the flare. This blueshift was consistent with plasma upflows driven by chromospheric evaporation or a prominence eruption. The corresponding plasma mass was estimated to be $\sim3.9\times10^{20}$~g for the evaporation scenario or $3.2\times10^{21}~\mathrm{g} < M{\mathrm{p}} < 8.8\times10^{21}$~g for the prominence eruption scenario.
A detailed analysis of this event is presented in \citet{Wang+etal+2026}.

\subsubsection{Recurrent Soft Transients in the Central Region of the Coma Cluster}\label{sec:Coma}

\begin{figure*}
        \centering
        \includegraphics[width=1.0\linewidth]{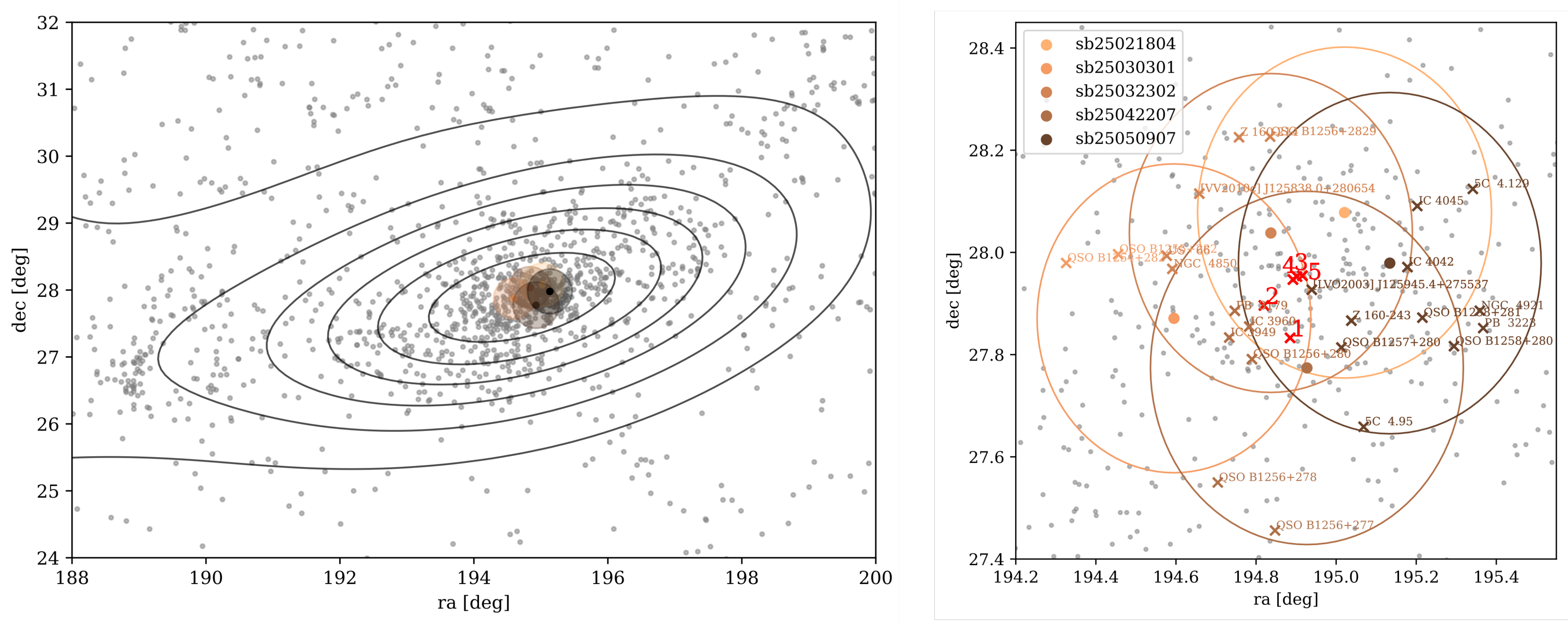}
        \caption{Left: positions of the five triggers in the Coma cluster, shown with density contours \citep{Mahajan2018}. Right: zoom on the region containing the five triggers, displaying the 90\ \% localization contours.}
        \label{fig:coma_triggers}
    \end{figure*}

Among the ECLAIRs triggers with unidentified origin, sb25021804, sb25030301, sb25032302, sb25042207, and sb25050907 are spatially coincident with the Coma cluster. These triggers occurred within a relatively short interval, on  February 8th, March 3rd, March 23rd, April 22nd, and May 9th 2025, respectively. Their sky positions, overlaid on the galaxy density contours of the Coma cluster, are shown in the left panel of Figure \ref{fig:coma_triggers} \citep{Mahajan2018}. The right panel of Figure \ref{fig:coma_triggers} presents a zoomed view of the region, including the 90\% localization confidence contours for each trigger (see \citealt{Goldwurm+etal+2026} and \citealt{Godet+etal+2026}). The red markers indicate sources lying within the five localization regions. All five triggers exhibit similar observational characteristics, detected onboard by the UGTS on a timescale of 1400~s and on the 5–8 keV energy band.

\begin{figure}
        \centering
        \includegraphics[width=1\linewidth]{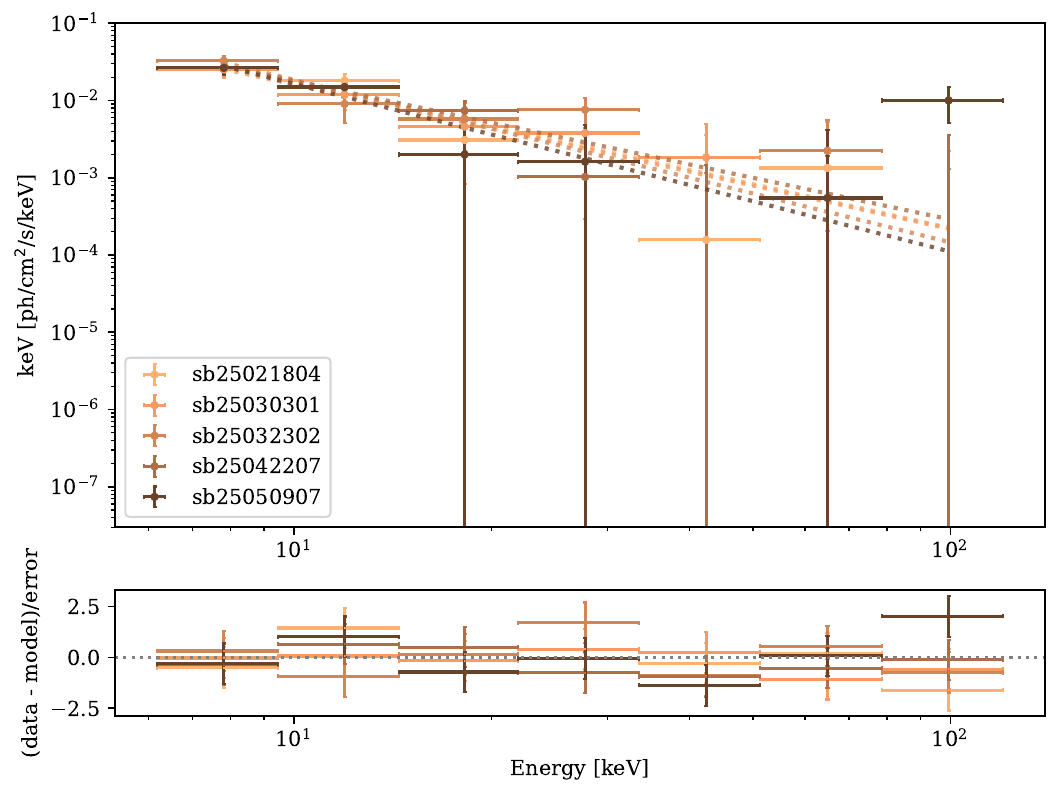}
        \caption{ECLAIRs hard X-ray spectrum of the five triggers in the Coma cluster, fitted with a power-law model.}
        \label{fig:coma_spectra}
    \end{figure}

To investigate whether these triggers originate from a common source, the ECLAIRs spectra were extracted for each event. Figure \ref{fig:coma_spectra} shows the resulting spectra fitted with a power-law model using XSPEC v12.14.1, and Table \ref{tab:coma_spectral_parameters} lists the corresponding best-fit parameters. The spectral analysis reveals that the five events are consistent with one another in both flux and spectral shape, suggesting a common origin. The inferred photon indices are notably soft, implying a possible thermal spectrum. We therefore also fitted the spectra with a blackbody model (\texttt{bbody} in XSPEC), obtaining characteristic temperatures $kT$ in the range $\sim 1$–3 keV (see Table \ref{tab:coma_spectral_parameters}). 

\begin{table*}[ht]
\centering
\begin{tabular}{l|ccc|ccc}
\hline
\hline
\textbf{Trigger name} & \multicolumn{3}{c|}{\textbf{Powerlaw model}} & \multicolumn{3}{c}{\textbf{Blackbody model}} \\
 & $\Gamma$ & Norm & $\chi^2$/dof & $kT$ & Norm & $\chi^2$/dof \\
 & & [ph/s/cm$^2$/keV] & [keV] &  & [$ \times 10^{36}$ ergs/s/10 kpc$^2$] & \\
\hline
sb25021804 & $2.48^{+0.45}_{-0.35}$ & $0.45^{+0.61}_{-0.24}$ & $8.23/6$ & $2.11^{+0.42}_{-0.36}$ & $5.07^{+1.16}_{-1.15}$ & $4.27/6$\\
sb25030301 & $2.67^{+0.61}_{-0.42}$ & $0.71^{+1.41}_{-0.42}$ & $1.95/6$ & $1.68^{+0.28}_{-0.25}$ & $6.27^{+0.86}_{-0.87}$ & $4.37/6$\\
sb25032302 & $2.78^{+0.50}_{-0.38}$ & $1.05^{+1.53}_{-0.56}$ & $6.05/6$ & $1.50^{+0.27}_{-0.24}$ & $7.85^{+1.35}_{-1.27}$ & $9.05/6$ \\
sb25042207 & $2.52^{+0.31}_{-0.26}$ & $0.53^{+0.43}_{-0.22}$ & $4.19/6$ & $1.96^{+0.31}_{-0.27}$ & $6.42^{+0.83}_{-0.83}$ & $7.04/6$\\
sb25050907 & $2.63^{+0.50}_{-0.39}$ & $0.60^{+0.91}_{-0.33}$ & $9.85/6$ & $1.90^{+0.39}_{-0.39}$ & $6.11^{+1.13}_{-1.13}$ & $6.63/6$ \\
\hline
\end{tabular}
\caption{Spectral parameters for the five triggers located in the central region of the Coma cluster, obtained using either a power-law or a blackbody model.}
\label{tab:coma_spectral_parameters}
\end{table*}

As shown on Figure \ref{fig:coma_triggers} (right panel), several known X-ray sources lie within the combined localization regions. We selected these sources using the SIMBAD database by applying a keyword-based classification, using the following keywords: \textit{X, BH, Gam, Neutron, NS, Psr, Magnetar, AGN, QSO, BLLAc, Sy1} and \textit{Sy2}. These keywords allow the identification of X-ray and gamma-ray emitters that are potentially detectable in the hard X-ray band. In addition, we excluded sources belonging to specific catalog (e.g., \textit{CXO, 2XMM, 3XMM, 4XMM, SDSS, 2MASS, PSZ2}, among others). These catalogues predominantly contain faint or extended sources whose expected flux is likely below the detection threshold of the ECLAIRs instrument. This exclusion criterion was applied to reduce contamination from sources unlikely to be detected and to focus the analysis on astrophysically relevant candidates for hard X-ray observations.

The sources located within the five error circles of these triggers are the quasar QSO~B1256+281, the active galactic nucleus NGC~4872, the ROSAT source RX~J125932.4+274958, the unidentified X-ray source 2E~1257.1+2813, and a candidate ultraluminous X-ray source (ULX), [SRW2012] Src. 6 \citep{Sutton2012}. QSO~B1256+281, NGC~4872, and RX~J125932.4+274958 are unlikely counterparts, since their reported X-ray fluxes cannot account for the observed ECLAIRs emission.
In addition, ULX flares are generally expected to exhibit relatively hard spectra \citep{Pintore2025, Ghosh2022}, making an association with [SRW2012] Src. 6 unlikely.

Still assuming that the triggers originate from the same source, the most plausible explanations are that the emission arises from 2E 1257.1+2813 or from a previously unidentified X-ray source within the Coma cluster region. However, the relatively low signal-to-noise ratio of the ECLAIRs detections limits our ability to draw definitive conclusions. Follow-up observations, combined with more detailed spectral and temporal analyses, are needed to reliably determine the nature of these recurrent triggers.

\subsubsection{Other unidentified ECLAIRs triggers}\label{sec:unidentified}

Beyond the five unidentified triggers consistent with the Coma cluster, an additional four unidentified ECLAIRs triggers were detected by 31 August 2025.

    \begin{figure*}
        \centering
      \includegraphics[width=\linewidth]{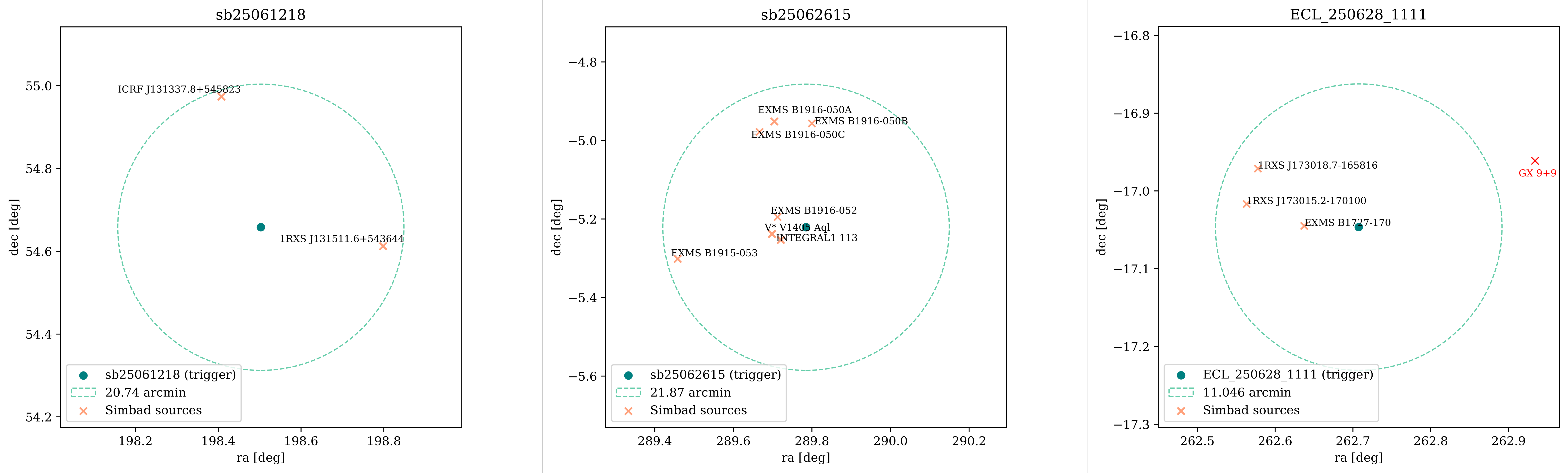}
        \caption{90\% localization error circles of the unidentified ECLAIRs triggers sb25061218, sb25062615 and ECL\_250628\_1111 from left to right. Catalogued X-ray sources present in the error box are indicated on the plots.}
        \label{fig:unknown_det}
    \end{figure*}
\begin{itemize}

\item \textit{sb25020104:} The event was detected by the UGTS/IMT (see Section \ref{sec:ugts}) on February 1st 2025 at 14:08:20.314\,UTC, in the 5--8\,keV energy band over a time window of 1310.72\,s. No known X-ray source is located within the localization error circle 14.36\,arcmin. No slew was performed, as the signal-to-noise ratio did not reach the threshold required to trigger a slew.\\

\item \textit{sb25062615:} The event was detected by the UGTS/IMT (see Section \ref{sec:ugts}) on June 25th 2025 at 04:16:53.871\,UTC in the 5--20 keV energy band over a time window of 40.96\,s seconds. Figure \ref{fig:unknown_det} (left panel) shows the trigger position together with the known X-ray sources located within the error circle. Five EXOSAT Medium Energy slew-survey catalog (EXMS) \citep{exmsCatalog} sources are consistent with the error region, but their catalogued fluxes are too low for the source to be detected by ECLAIRs. The most likely candidates are therefore V* V1405 Aql (4U 1916-05), a LMXB \citep{Krivonos2022}, or INTEGRAL1 113, which is also classified as an LMXB \citep{Bird2004}.\\

    
\item \textit{sb25061218 / SVOM~J13140+5439:} The event was detected by the UGTS/IMT (see Sect. \ref{sec:ugts}) on June 12th 2025 at T$_0$~=~20:34:38.714~UTC, in the 8–120 keV energy band over a timescale of 81.92 s. Figure \ref{fig:unknown_det} (middle panel) shows the trigger position together with the known X-ray sources within the error circle. Two sources are present, ICRF~J131337.8+545823, a Seyfert galaxy, and the ROSAT source 1RXS~G115.50+62.16, both of which are unlikely counterparts due to their faint recorded X-ray fluxes. In addition, Swift observed the field between T$_0$+5.9~ks and T$_0$+7.7~ks, detecting a source consistent with the quasar [VV2006] J131446.6+544804 (z = 0.48968), also observed with the Observatoire de Haute-Provence (OHP) T120 telescope at $r \simeq  18.3$~mag, about 0.3~mag brighter than catalogued values \citep{Adami2025}. This points to a possible association of the SVOM trigger with a phase of enhanced multi-wavelength quasar activity. However, the large discrepancy between the X-ray fluxes measured by SVOM and Swift leaves open the possibility that the event is either a new unrelated transient source (SVOM~J13140+5439) or a GRB coincident with the quasar flare. This trigger was reported in an Astronomer’s Telegram \citep{Rodriguez2025}.\\

\item \textit{ECL\_250628\_1111:} On June 28th, 2025, at 11:11:26.60~UTC, a X-ray burst was detected in the 4–20 keV energy range using the OFT (see Section \ref{sec:oft}). The burst lasted approximately 10~s and exhibited characteristics typical of a Type I thermonuclear burst from a LMXB. Figure \ref{fig:unknown_det} (right panel) shows the trigger position with its associated localization uncertainty, along with the X-ray sources present in the field of view. Two ROSAT sources, 1RXS J173015.2-170100 and 1RXS J173018.7-165816, as well as the EXOSAT source EXMS B1727-170, lie within the error circle; however, the burst is unlikely to have originated from any of them, as their catalogued fluxes are too low for these sources to be detected by ECLAIRs. The known source GX 9+9 is located 13.85~arcmin from the trigger position, outside the 90\% confidence level error circle. Therefore, even if it cannot be totally ruled out, it is likely that this X-ray burst does not originate from GX~9+9. This suggests that it may have originated from a previously unknown source, tentatively designated as SVOM J17308-1702. A preliminary report of this detection was published in an Astronomer’s Telegram \citep{Guillot2025}.

\end{itemize}

\section{SVOM Target-of-Opportunity Follow-Up of External Alerts}

In addition to the GP pointed and planned observations and serendipitous high-energy source detections by ECLAIRs, SVOM conducts a Target of Opportunity (ToO) program. GP-ToO observations can be requested annually through the internal CfP, for known targets exhibiting unpredictable active phases (see Section \ref{sec:GP_description}). Beyond these planned requests, SVOM also triggers unanticipated ToOs throughout the year, including observations of newly discovered targets or follow-ups of (catalogued) sources detected by other facilities. In the following section, we present preliminary results from the SVOM ToO program, focusing on follow-up observations of sources detected by Einstein Probe as well as multi-messenger ToOs.

\subsection{Follow-Up of soft X-ray sources detected by Einstein Probe} 

SVOM and Einstein Probe \citep{2022hxga.book...86Y} are complementary missions in time‑domain astrophysics. Einstein Probe's wide‑field soft X‑ray survey enables discovery and high‑cadence monitoring of diverse transient and variable sources, while SVOM specializes in detecting hard X-ray transients and performing rapid multi‑wavelength follow‑up in X‑ray and optical bands. Together, they provide broad spectral coverage from soft to hard X‑rays and cover both discovery and characterization roles. Regarding Observatory Science topics, SVOM’s strengths lie in its multi‑wavelength capabilities, whereas Einstein Probe excels at wide-field monitoring and sensitivity to faint soft X‑ray outbursts as generally observed in AGNs for instance. Working in synergy, the two missions may enhance the monitoring and understanding of variable high-energy sources across the sky from 0.5 up to 150\,keV.

To date, SVOM has followed up an increasing number of sources discovered by Einstein Probe, including EP240904a, EP250305a, and EP250623a, as well as monitored X-ray binaries, highlighting the synergy between the two missions. 

\begin{itemize}



\item \textit{GX~339--4:} GX~339--4 is a well-known black hole binary that frequently undergoes outbursts. In 2025, it experienced a very faint outburst, initially detected in the optical band and subsequently confirmed through Einstein Probe/FXT follow-up observations. During this event, GX~339--4 exhibited its lowest QPO frequency ever recorded. The photon index remained nearly constant with time and flux, indicating the absence of a state transition and confirming this as another failed outburst. Coordinated observations by SVOM/MXT and Einstein Probe/FXT enabled precise determination of the source’s spectral parameters and flux evolution \citep{Zhao_2026}.  \\

\item \textit{EP250702a / GRB~250702B,D,E:} 
EP250702a (also reported as GRB250702B/D/E) is an unusual high-energy transient detected on July 2nd 2025, characterized by several gamma-ray bursts spanning several hours and detected by Fermi/GBM \citep{GBM_GRB250702}, Konus-WIND \citep{Konus_GRB250702}, Swift/BAT \citep{BAT_GRB250702}, Insight-HXMT and GECAM \citep{2026ApJ...997L..45Z}. These events may be associated with the soft X-ray transient EP250702a detected by Einstein Probe \citep{2025GCN.40906....1C} and independently by MAXI/GSC \citep{MAXI_GRB250702}.
Follow-up observations revealed a rapidly fading, extremely red ($H-K  = 1.42\,\pm\,0.06$\, AB\,mag) infrared counterpart associated with an extragalactic host galaxy, while optical emission was undetected, suggesting either strong dust extinction or intrinsically red emission \citep{2025arXiv250714286L}. The prolonged, multi-episode gamma-ray activity and late-time X-ray behavior have motivated several competing interpretations. One leading scenario invokes an ultra-long GRB produced by the collapse of a supergiant star, with sustained fallback accretion onto a newly formed compact object powering prolonged jet activity \citep{2026ApJ...997L..45Z}. Alternatively, the event may represent a tidal disruption event (TDE), potentially involving an intermediate-mass black hole and the disruption of a white dwarf, capable of producing extended high-energy emission and episodic variability \citep{2025arXiv250925877L}. In this context, reported soft X-ray decay slopes consistent with fallback accretion ($\propto t^{-5/3}$) and possible quasi-periodic features may been cited as supporting evidence for a TDE-like origin with some caveats \citep{2026ApJ...997L..45Z}, although current astrometric constraints do not yet robustly establish a nuclear origin.
SVOM/GRM detected hard X-ray activity using event-by-event data retrieved via the X-band ground station, consistent with contemporaneous detections by other facilities (\citealt{2025GCN.40906....1C} and Figure \ref{fig:GRM_250702}). At the time of the hard X-ray triggers, the refined position reported by Einstein Probe/WXT (R.A.~=~284.6911$^\circ$; Dec = –7.8715$^\circ$, J2000; GCN~40906) was $\sim$48$^\circ$ from the SVOM optical axis, placing it just outside the ECLAIRs FoV. SVOM ToO observations were conducted on  July 3rd and 5th 2025. No significant hard X-ray emission was detected by ECLAIRs. A preliminary upper limit on the 4--120\,keV flux of $\sim~1.3\times10^{-9}$~\,erg/s/cm$^2$ was therefore derived assuming a power-law spectrum with photon index $\Gamma=2.0$. SVOM/VT also reported non-detections on the same days, providing AB magnitude upper limits in the VT B and R bands of 23.8 and 23.6, respectively \citep{2025GCN.41122....1L}.

\begin{figure}
    \centering
    \includegraphics[width=\linewidth]{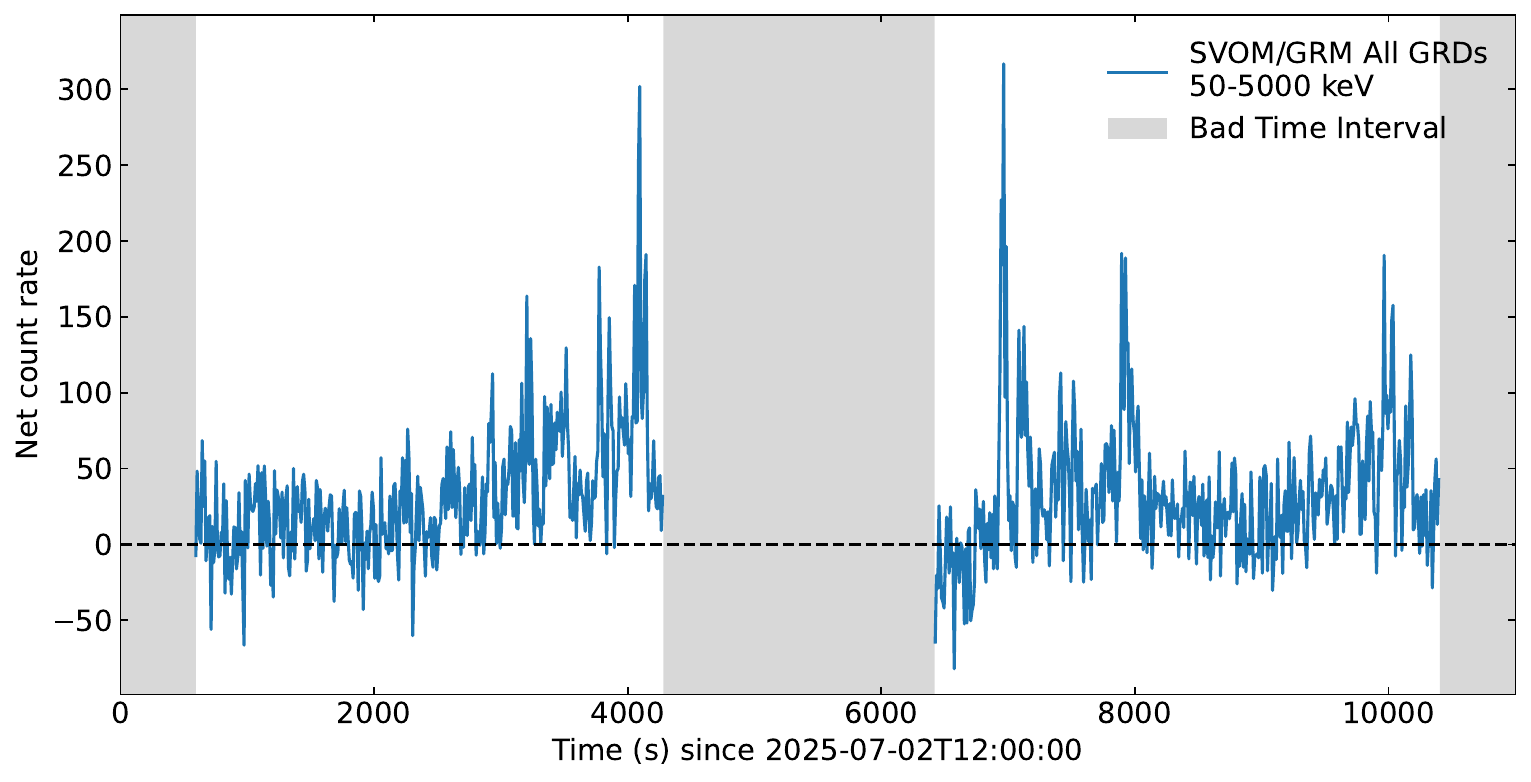}
    \caption{GRM lightcurve showing the detection of GRB\,250702D and B respectively around 13:09:02 UTC and 13:56:05 on July 2nd 2025. The gray shaded regions indicate periods of South Atlantic Anomaly (SAA) passage and Earth occultation.}
    \label{fig:GRM_250702}
\end{figure}

\end{itemize}

\subsection{Multi-Messenger Follow-Up of the KM3NeT Neutrino Event KM3-230213A} 
\begin{figure}
        \centering
                \includegraphics[width=1\linewidth]{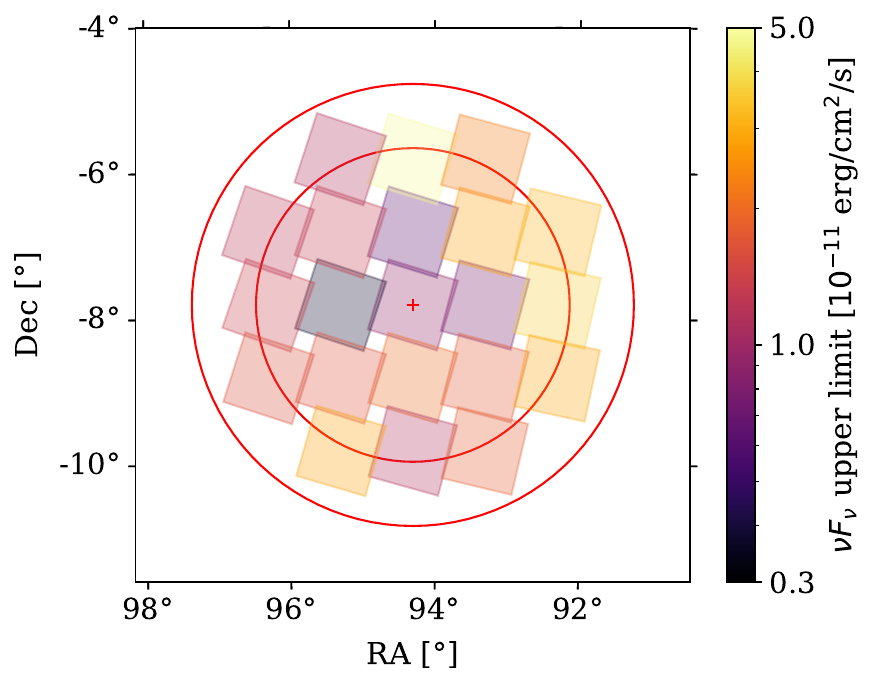}
                        \includegraphics[width=1\linewidth]{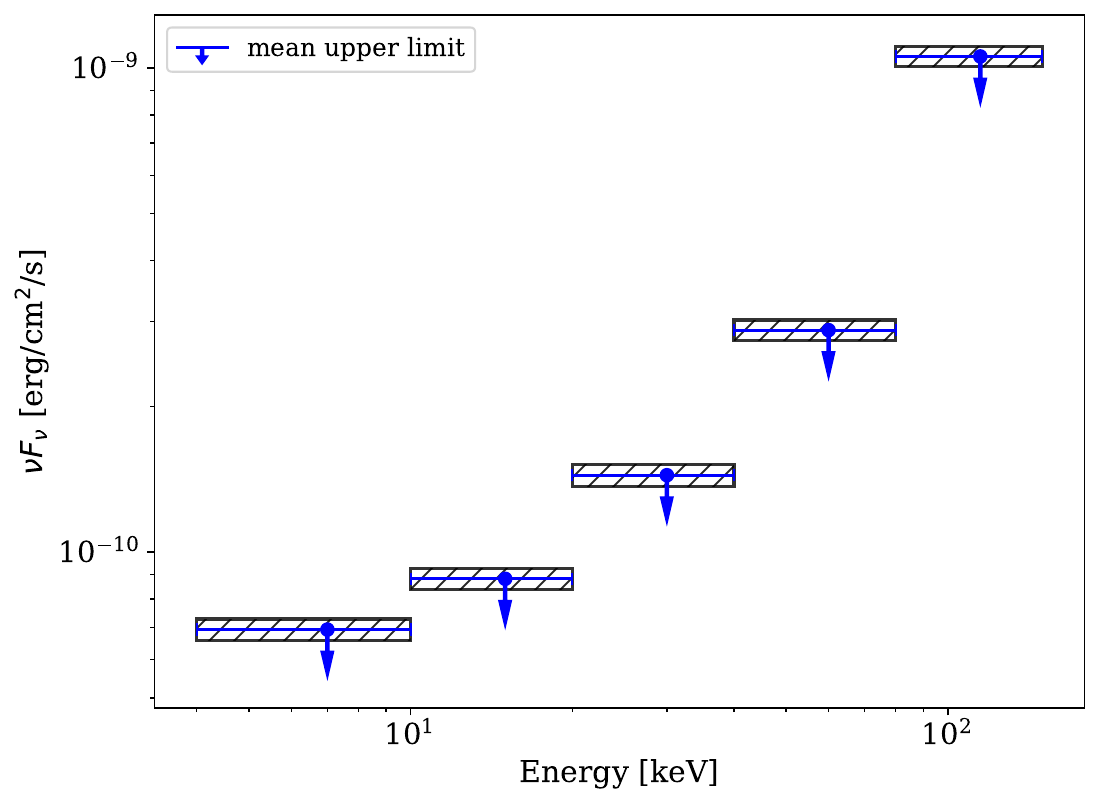}
        \caption{Upper panel: 3$\sigma$ upper limits derived from the tiling of the KM3-230213A neutrino error region with MXT over the 0.5--10\,keV energy band. The concentric red circles indicate the 90\% and 99\% neutrino error regions. Lower panel: ECLAIRs 3$\sigma$ upper limits on the energy flux on different energy ranges for blazars in the neutrino error box. Adapted from \citet{Lincetto+etal+2026}.}
        \label{fig:ToO_KM3}
    \end{figure}

During SVOM’s commissioning phase, we conducted a comprehensive end-to-end test of the multi-messenger ToO system, designed to validate both the triggering infrastructure and the automated tiling and execution procedures. In addition, the mission carried out its first Multi-Messenger ToO program (ToO-MM) from December 16th to 17th, 2024, targeting the 90\% confidence-level neutrino localization region associated with the ultra-high-energy event KM3-230213A detected by the KM3NeT neutrino telescope \citep{2025Natur.638..376K} in 2023. Although not performed in near real time, this observation provided a full validation of the operational chain, from alert ingestion to automated tiling definition and observation execution, thereby demonstrating the readiness of SVOM for multi-messenger follow-up. The MXT instrument performed 21 tiled pointings with an average exposure of $\sim$1200~s. Data were processed through the dedicated pipeline, yielding no detected soft X-ray sources. Consequently, 3$\sigma$ flux upper limits were derived for each tile, accounting for exposure variations, and are represented spatially in Figure \ref{fig:ToO_KM3} (upper panel). These limits provide useful constraints on soft X-ray emission from possible neutrino emitters in the error box.

Simultaneously, ECLAIRs monitored the region in the 4–150 keV range for a total effective exposure of 25.6 ks. No hard X-ray emission was detected. Using variance maps produced by the ECPI pipeline \citep{Goldwurm+etal+2026}  and assuming a power-law spectrum with $\Gamma$ = 2, 3$\sigma$ upper limits were derived on the full energy band of the instrument (see Fig. \ref{fig:ToO_KM3}, lower panel). These limits complement the MXT survey, effectively ruling out bright hard X-ray activity during the ToO window.

Overall, these SVOM observations represent the first ToO-MM conducted by the mission, providing an important opportunity to test both the system and the tiling strategy defined for multi-messenger follow-up. It serves as a demonstration of SVOM’s capability to conduct coordinated optical and X-ray observations in response to neutrino events \citep{Lincetto+etal+2026}. 

\section{Conclusions and Perspectives}
\label{sect:conclusion}

The first year of SVOM operations demonstrated the mission’s strong potential for Observatory Science, extending well beyond its primary GRB detection and characterisation objectives.  The large FoV and continuous monitoring of the sky provided by ECLAIRs have enabled the detection of hundreds of non-GRB triggers, including X-ray binaries, blazars, stellar flares, and unidentified events. These early results position SVOM as a key contributor to time-domain high-energy astrophysics.

The observations presented in this paper highlight several notable achievements. The detection of burst oscillations in the Type I burst of 4U~0614+091 demonstrates ECLAIRs’ sensitivity and timing capabilities for probing neutron-star surface phenomena. The first blazar outburst detected by SVOM, from 1ES~1959+650, illustrates the mission’s ability to capture broadband variability and to trigger coordinated multiwavelength follow-up observations. In addition, the detection of chromospheric evaporation in HD~22468 reveals SVOM’s unexpected reach into stellar physics, while the monitoring of Aql~X-1 and recurrent soft transients in the Coma cluster region underscores its capacity to track spectral evolution and identify rare or previously unknown sources.

Looking ahead, refined calibration of the instruments will continue, and the capabilities of the ECLAIRs Quick-Look Analysis and the Offline Trigger will be further expanded, improving sensitivity to faint and short-duration events and enabling more robust source detection and classification. In parallel, consolidated scientific products—such as light curves and spectra of non-GRB sources and triggers are planned to be made publicly available through dedicated multi-wavelength catalogues, broadening the mission’s legacy value to the wider astrophysical community. Moreover, increased synergy and coordination with other facilities, including ground-based optical and radio observatories, will further enhance the scientific return. SVOM’s ability to rapidly respond to external alerts also positions it as a key asset for multi-messenger astronomy, particularly in the context of neutrino and gravitational-wave triggers. 

As SVOM enters its mature operational phase, the Observatory Science program is poised to deliver deeper insights into accretion physics, particle acceleration, magnetospheric processes, explosive stellar phenomena, etc. A Guest Program (GP) call for proposals will be released in Spring 2026, and researchers outside the SVOM collaborations who are interested in applying for SVOM observing time are encouraged to get in touch with a SVOM co-investigator\footnote{The full list of SVOM co-Investigators and Affiliate Scientists can be found at \url{https://fsc.svom.org/home/collaboration/collaborators}}. With its broad discovery space and growing data volume, SVOM promises a rich and diverse scientific harvest in the years to come.

\begin{acknowledgements}
The Space-based multi-band astronomical Variable Objects Monitor (SVOM) is a joint Chinese-French mission led by the Chinese National Space Administration (CNSA), the French Space Agency (CNES), and the Chinese Academy of Sciences (CAS). We gratefully acknowledge the unwavering support of NSSC, IAMCAS, XIOPM, NAOC, IHEP, CNES, CEA, and CNRS. 
\end{acknowledgements}

\bibliographystyle{raa}
\bibliography{bibtex}

\label{lastpage}

\end{document}